\documentclass[english,aps,prb,floatfix,superscriptaddress,reprint]{revtex4-1} 
\usepackage[utf8]{inputenc}
\usepackage{graphicx}
\usepackage{amsmath,amsfonts,amssymb}
\usepackage{color}
\usepackage{pdfpages}
\usepackage[capitalize]{cleveref}
\usepackage{braket}
\usepackage{xcolor}
\usepackage{ulem}

\crefname{appsec}{Appendix}{Appendices}

\newcommand{\nairo}{Na\ensuremath{_2}IrO\ensuremath{_3}}
\newcommand{\rucl}{\mbox{\ensuremath{\alpha}-RuCl\ensuremath{_3}}}

\newcommand{\fieldterm}{\ensuremath{H_\text{f}}}

\newcommand{\hrucl}{\ensuremath{H_{\text{RuCl}}}}

\newcommand{\imag}{{\mathrm{i}\mkern1mu}}
\newcommand{\e}{{\mathrm{e}\mkern1mu}}

\newcommand{\transposedSign}[0]{\intercal}
\newcommand{\transposed}[0]{^\transposedSign}

\newcommand{\phiFM}{\ensuremath{\phi_{\text{tet}}}}

\newcommand{\RIXS}{\ensuremath{\mathcal{I}_{\text X}}}

\makeatletter
\AtBeginDocument{\let\LS@rot\@undefined}
\makeatother

\begin{document}

\title{Kitaev honeycomb models in magnetic fields: \protect\linebreak Dynamical response and hidden symmetries}
\date{\today}

\begin{abstract}
Motivated by recent reports of a
field-induced {\it intermediate phase} (IP) in the antiferromagnetic honeycomb Kitaev model
that may be a spin liquid 
 whose nature is distinct from the Kitaev $\mathbb Z_2$ phase,
we present a detailed numerical study on
 the nature and dynamical response (such as 
dynamical spin-structure factors and resonant inelastic x-ray scattering intensities)  
of this field-induced IP and neighboring
phases in a family of Kitaev-based models related by hidden symmetries and duality
transformations. We further
 show that the same field-induced IP can appear in models relevant for
 \rucl, which exhibit a ferromagnetic Kitaev coupling and additional interactions. 
In \rucl, the IP represents a new phase, that is likely independent from the 
 putative field-induced (spin-liquid) 
phase recently reported from thermal Hall conductivity measurements. 
\end{abstract}

\author{David A. S. Kaib} 
\email{\href{mailto:kaib@itp.uni-frankfurt.de}{kaib@itp.uni-frankfurt.de}}
\affiliation{Institut f\"ur Theoretische Physik, Goethe-Universit\"at Frankfurt,
Max-von-Laue-Strasse 1, 60438 Frankfurt am Main, Germany}

\author{Stephen M. Winter}
\affiliation{Institut f\"ur Theoretische Physik, Goethe-Universit\"at Frankfurt,
Max-von-Laue-Strasse 1, 60438 Frankfurt am Main, Germany}

\author{Roser Valent{\'\i}}
\affiliation{Institut f\"ur Theoretische Physik, Goethe-Universit\"at Frankfurt,
Max-von-Laue-Strasse 1, 60438 Frankfurt am Main, Germany}

\maketitle

\section{Introduction}
In magnets with strongly frustrated interactions, quantum spin liquids 
 can arise as exotic phases of matter that lack spontaneous symmetry-breaking down to zero temperature and feature long-range entanglement and fractionalized excitations. 
A prime example is the exactly solvable Kitaev 
model on the honeycomb lattice~\cite{kitaev2006anyons}, which hosts a 
quantum spin liquid ground state of itinerant Majorana fermions
 that couple to a static $\mathbb Z_2$ gauge field. 

Material realizations of the Kitaev model have been heavily sought after, and a
mechanism\cite{jackeli2009mott} relying on an intricate interplay of strong
electronic correlations, crystal field splitting and spin-orbit coupling has
brought candidate materials such as \nairo, various polymorphs
of Li$_2$IrO$_3$, and \rucl\ to the
forefront of research. However, these materials all display magnetic order at
low temperatures, which is a result of additional magnetic couplings extending
beyond the pure Kitaev coupling. 
 With the goal of unravelling potential residual fractionalized excitations
reminiscent of the pure Kitaev model in materials,
 there have been various routes to suppress the magnetic order, including the
application of pressure\cite{cui2017high,biesner2018detuning,bastien2018pressure,wang2018pressure}, finite temperatures\cite{sandilands2015scattering,nasu2016fermionic,do2017majorana}
 or a magnetic field\cite{johnson2015monoclinic,yadav2016kitaev,sears2017phase,wolter2017field,baek2017observation,wang2017magnetic,zheng2017gapless,ponomaryov2017unconventional,banerjee2018excitations,hentrich2018unusual,kasahara2018half,janssen2019heisenberg}. For the
latter case, a putative field-induced 
phase\cite{baek2017observation,wang2017magnetic,zheng2017gapless,banerjee2018excitations,kasahara2018half}
in \rucl, that lacks magnetic order, is under strong scrutiny. Here, a
half-integer quantized thermal Hall conductivity was 
recently reported\cite{kasahara2018half} for fields tilted by 
$30^\circ$ and $45^\circ$ out of the honeycomb plane.
Such measurements have motivated much theoretical effort to analyze both the
original Kitaev model as well as models with realistic extended interactions in
magnetic fields. 

On the theoretical side, it is known that
without a magnetic field, the pure ferromagnetic (FM) and antiferromagnetic
(AFM) versions of the Kitaev model are related by a unitary transformation and
thus share the same topological properties. For both coupling signs, the
Kitaev spin liquid (KSL) survives under a weak magnetic field, where it becomes
gapped and hosts non-Abelian anyonic excitations\cite{kitaev2006anyons}. 
The effects of a stronger field after suppressing the KSL state
 have however recently gained much theoretical interest due to the
discovery of a field-induced {\it intermediate phase} (IP) 
 in the AFM model\cite{zhu2018robust,gohlke2018dynamical}. 
This phase is separated from the low-field KSL and the high-field polarized state by phase transitions and could itself be a quantum spin liquid \cite{jiang2018field,zou2018field,patel2018magnetic,hickey2019emergence,jiang2019tuning}.

Motivated by these findings, 
(i) we perform a detailed analysis
of the nature and dynamical response
of the field-induced {\it intermediate phase}   and neighboring
phases in a family of Kitaev-based models related by hidden symmetries and duality
transformations and, (ii) we investigate
 the relevance of the IP phase for real materials, in particular for
 \rucl, and discuss the relation of this phase to the putative
field-induced phase reported from thermal Hall conductivity 
measurements~\cite{kasahara2018half}.    

The paper is organized as follows;
 in \cref{sec:Kitaev_uni_field} we 
revise the 
properties of the Kitaev model and present numerical results
 for various dynamical response functions of
the FM and AFM Kitaev model in uniform magnetic fields. This includes the
dynamical spin-structure factor, which can be accessed by e.g.\ inelastic
neutron scattering (INS) or electron spin resonance (ESR) experiments, and
 dynamical bond correlations, that contribute to resonant inelastic x-ray
scattering (RIXS) and Raman scattering. Furthermore, we probe directly static
and dynamic flux-flux correlations that appear under field. In
\cref{sec:general_fieldsA}, by making use of duality transformations
in the Kitaev model, we study the effect  of generalized
(non-collinear) magnetic fields.
 Numerically, we find that the field-induced
IP of the AFM Kitaev model is highly unstable against certain
non-uniform field rotations, and could manifest as a line of critical points in the
parameter space of such generalized fields. 
In \cref{sec:extendedinteractions} we discuss the relevance of the
 field-induced IP phase of the AFM Kitaev model for real materials.
 By utilizing hidden symmetries in the parameter space of extended interactions,
 we show that the same IP phase can appear in realistic models for \rucl,
 that possess a \textit{ferromagnetic} Kitaev coupling
 and additional interaction
terms, for fields perpendicular to the honeycomb plane.

\section{Pure Kitaev Model\protect\linebreak in Uniform Magnetic Fields  
\label{sec:Kitaev_uni_field}} 
\subsection{Introduction and definitions}

 \begin{figure}
\includegraphics[width=1\linewidth]{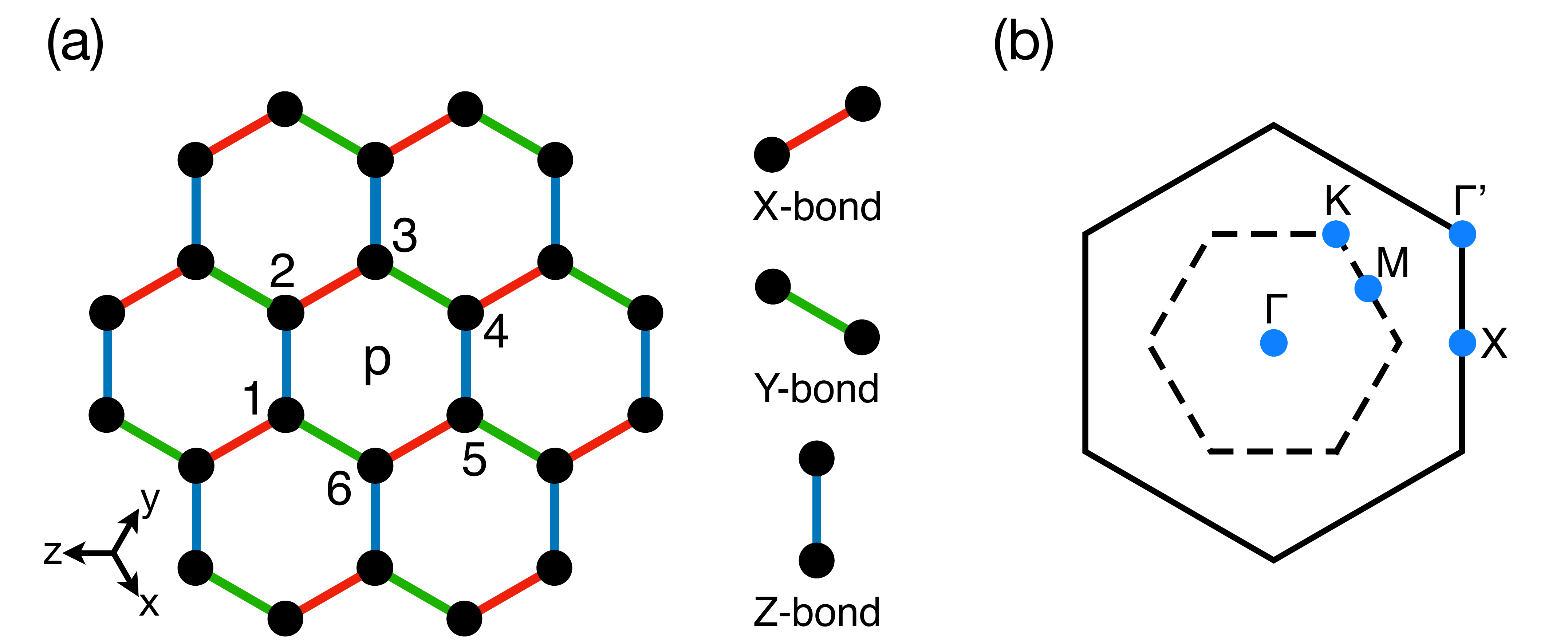}
\caption{(a)~Honeycomb lattice with definitions of cubic axes $x,y,z$ and bond types $\mathrm X, \mathrm Y, \mathrm Z$. 
 (b)~High-symmetry $k$-points in the first (dashed) and third Brillouin zone (solid border).
}
\label{fig:lattice}
\end{figure}

The Kitaev honeycomb model\cite{kitaev2006anyons} is defined by a Hamiltonian consisting of bond-dependent Ising terms:
\begin{equation}
H_{K}=K\sum_{\braket{ij}_\gamma} S_i^\gamma S_j^\gamma,
\label{eq:Hkitaev}
\end{equation}
where $\gamma\in\{x,y,z\}$ corresponds to the type of bond that connects $i$ and $j$, according to \cref{fig:lattice}(a).

  The classical version of the model exhibits an extensive
 degeneracy\cite{baskaran2008spin}, 
 that is lifted in the quantum model in favor of the KSL ground states. The latter can be found 
exactly \cite{kitaev2006anyons} by representing the spins through fermionic Majorana operators $\{b_i^x, b_i^y, b_i^z, c_i\}$ as $S_i^\gamma=\frac\imag 2 b_i^\gamma c_i$. This representation introduces a 
  $\mathbb{Z}_2$ 
 gauge redundancy; physical states are spanned by unique configurations of \textit{matter} fermions, and $\mathbb{Z}_2$ \textit{flux} degrees of freedom. The latter are governed by Wilson loop operators $\{ W_L\}$, which commute with $H_K$. The shortest of such loops are associated with hexagonal plaquette operators ${W}_p=2^6 S_1^x S_2^y S_3^z S_4^x S_5^y S_6^z$, where the site indices refer to those in \cref{fig:lattice}(a). 
 In the ground states,
the flux density 
\begin{equation}
  \braket{n_p} \equiv \tfrac12 \left(1-\braket{W_p}\right)
  \label{eq:flux_density}
\end{equation}
vanishes and the model can be written in terms of a free-fermion problem.

The exact solubility has facilitated a deep understanding of various response functions. 
 The application of a single spin operator $S_i^\mu$ 
 creates fluxes on two plaquettes, 
  in addition to excitations in the matter sector. As a result, the dynamical spin-structure factor 
  \begin{align}
\mathcal{S}(\mathbf{k},\omega) =\sum_{\mu=x,y,z} \int dt \ \e^{-\imag\omega t} \Braket{ S^\mu_{-\mathbf{k}}(t)\, S^\mu_{\mathbf{k}}(0)}
\end{align}
probes exclusively fluxful excitations, which form a continuum that is both dispersionless\cite{baskaran2007exact} and gapped, with intensity only above the two-flux gap\cite{knolle2014dynamics,knolle2015dynamics} $\Delta_\text{f} \simeq 0.065 |K|$.

 The dynamics of the bond operators $B_{ij}:=S_i^\gamma
S_j^\gamma$ and of the flux operators $W_p$ provide further intrinsic signatures
of the KSL and its field-induced phases.
We consider for the bond operators the correlation function 
\begin{align}
	\RIXS(\mathbf k, \omega) :=& \int dt\ \e^{-\imag \omega t} \Braket{ B_{-\mathbf k}(t)\, B_{\mathbf k}(0) }, \label{eq:RIXSintensity}  
	\\
	B_{\mathbf k}  :=& \frac1N \sum_i \e^{\imag \mathbf k \cdot \mathbf r_i} \sum_{\gamma={x,y,z}} S_i^\gamma S_{\gamma(i)}^\gamma, \label{eq:Bk_def}
\end{align}
where $\gamma(i)$ is the nearest neighbor of $i$ along a $\gamma$-bond and $N$ is the number of sites. 
Due to $[B_\mathbf{k},W_p]=0$, $\RIXS(\mathbf k,\omega)$ probes only dispersive fermionic matter excitations in the flux-free sector\cite{halasz2016resonant}, revealing gapless modes (with vanishing intensity) at $\mathbf k=\Gamma$ and $\mathbf k=\text K$ that reflect the Dirac spectrum of the underlying spinons\cite{kitaev2006anyons}. 
$\RIXS(\mathbf k,\omega)$ constitutes the main contribution to the spin-conserving channel of the resonant inelastic x-ray scattering (RIXS) intensity\cite{halasz2016resonant}. For $\mathbf k=\Gamma$, $\RIXS(\Gamma, \omega)$ also contributes to Raman scattering\cite{knolle2014raman}. 

Finally, we define the dynamical flux-structure factor
\begin{align}
	\mathcal W(\mathbf k, \omega) := & \ \int dt \ \e^{-\imag\omega t} \Braket{  W_{-\mathbf{k}}(t)\,  W_{\mathbf{k}}(0)} \label{eq:Wkw} ,\\
	   W_\mathbf k := & \  \frac{1}{N/2} \sum_{p}^{N/2}\e^{\imag\mathbf k \mathbf r_p}  W_p. 
  \label{eq:Wk}
\end{align}
 In the pure Kitaev model, $\mathcal W(\mathbf{k},\omega)$ has no intensity at finite $\omega$, reflecting 
 that fluxes are completely static. However, generic perturbations to $H_K$ may lead to dynamical fluxes. 
In this section, we focus on the effect of a uniform magnetic field: 
\begin{align}
  H = H_K - \sum_i \mathbf h \cdot \mathbf S_i \label{eq:Hkitaev_uniformfield}
\end{align}
For weak fields, Kitaev found using perturbative arguments, that the field induces a gap in the matter fermion spectrum, which then carries a nonzero Chern number\cite{kitaev2006anyons}.

The fate of the model beyond this perturbative weak-field limit has been the subject of much recent interest. 
The FM ($K<0$) and AFM ($K>0$) versions of the Kitaev model display rather different behavior under field, which can already be anticipated on the classical level:  In the case of FM coupling, a finite field instantly selects the polarized state out of the classically degenerate spin configurations\cite{janssen2016honeycomb}, suggesting a critical field of $h_\text{c}^\text{FM}=0$. For AFM coupling, the classical degeneracy
 is retained\cite{janssen2016honeycomb}  
 up to a field of $h_\text{c}^\text{AFM}=K$, as the polarized state does not fulfill the AFM spin-spin correlations preferred by the coupling.  
 For the quantum Kitaev models (FM and AFM), the stability of the zero-field KSL is determined instead by the flux gap $\Delta_\text{f}$, which provides an additional emergent low energy scale. Independent of the sign of $K$, excitations carrying finite flux acquire a dispersion on the order of $h$. Suppression of the zero-field spin liquid likely occurs when this dispersion exceeds the flux gap, leading to a proliferation of fluxes at a critical field strength of $h_\text{c} \sim \Delta_\text{f}$.

This implies a finite stability of the KSL in the FM Kitaev model. Indeed, various numerical studies\cite{jiang2011possible,zhu2018robust,gohlke2018dynamical,hickey2019emergence} have indicated a single phase transition as a function of field strength at $h_{\text{c}}^{\text{FM}}\approx 0.03|K|$, qualitatively independent of field direction\cite{hickey2019emergence}. The transition occurs directly to a {\it quantum paramagnet} (QPM) phase, that is smoothly connected to the {\it fully polarized state} of the ${h}\rightarrow\infty$ limit.

For the AFM model, the fact that $\Delta_\text{f} \ll K$ provides the possibility of an intermediate field regime where neither the KSL nor the polarized phase are the ground state. Consistently, recent studies have found an additional
{\it intermediate phase}\cite{zhu2018robust,gohlke2018dynamical,hickey2019emergence,jiang2018field,zou2018field,patel2018magnetic} (IP) for fields along the cubic $[111]$ direction (i.e.\ the direction perpendicular to the honeycomb plane) in the range $h_{\text{c}1}^\text{AFM} < h < h_{\text{c}2}^\text{AFM}$, with $h_{\text{c}1}^{\text{AFM}}\approx 0.4K$ and $h_{\text{c}2}^{\text{AFM}}\approx 0.6K$. The IP is thought to be gapless in the thermodynamic limit\cite{zhu2018robust,gohlke2018dynamical,hickey2019emergence,jiang2018field,zou2018field,patel2018magnetic}. Consequently, descriptions of the IP in terms 
of various types of spin liquids have been
proposed\cite{hickey2019emergence,jiang2018field,zou2018field,patel2018magnetic,jiang2019tuning}.
Interestingly, the stability of the IP also appears to depend on the
orientation of the field\cite{hickey2019emergence}; for fields along $[001]$ in
the AFM Kitaev model, Majorana mean-field
studies\cite{liang2018intermediate,nasu2018successive} found a possibly
different field-induced phase. 
In what follows, we focus on the IP that is  induced for
field directions including $[111]$.

\subsection{Ferromagnetic Kitaev model}

In order to study the phenomenology of the 
 transition between the KSL and the field-polarized phase, we first consider the FM Kitaev model ($K < 0$) in a uniform magnetic field $\mathbf{h} \parallel [111]$, described by \cref{eq:Hkitaev_uniformfield}. 
All shown results were obtained by exact diagonalization (ED) of the model on the 24-site cluster with periodic boundary conditions
shown in \cref{fig:lattice}(a). Other studied field directions (not shown) including $[1\bar10], [11\bar2],$ and $[001]$ yield qualitatively similar response. 

\begin{figure}
\includegraphics[width=\linewidth]{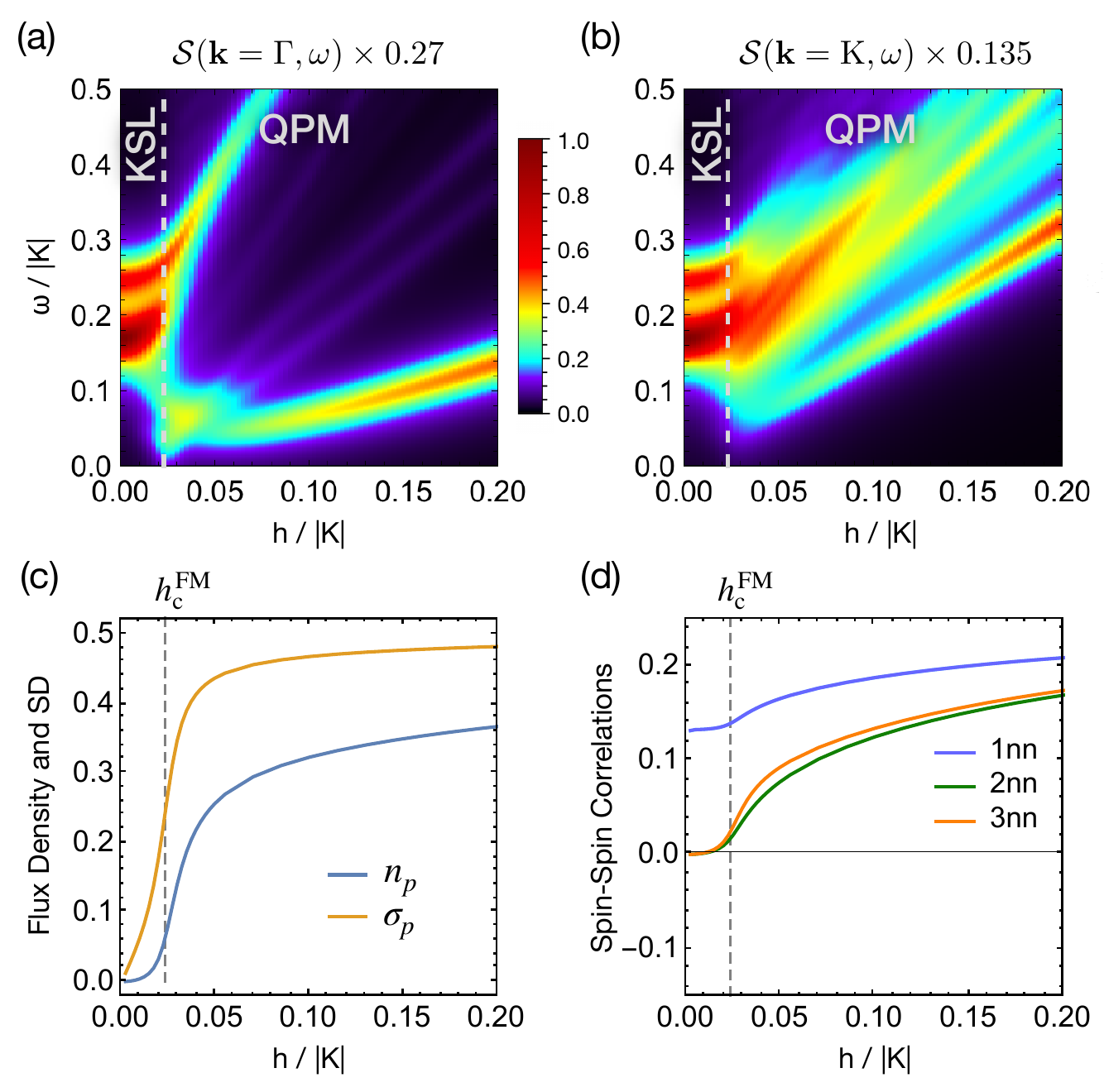}
\caption{
FM Kitaev model under $[111]$-field. (a,b)~Low-frequency dynamical spin-structure factor at selected $k$-points. Poles at $\omega=0$ are not shown.  (c)~Flux density $\braket{n_p}$  and standard deviation (SD) of the local flux density $\sigma_p=\sqrt{\braket{{n_p}^2}-\braket{n_p}^2}$. (d)~Static spin-spin correlations in real space. 1nn, 2nn, 3nn denote $\braket{\mathbf S_i \cdot \mathbf S_j}$ on first-, second- and third-nearest neighbors, respectively. 
 }
\label{fig:FM_DSF}
\end{figure}

 The evolution of the dynamical spin-structure factor  $\mathcal{S}(\mathbf{k},\omega)$ is shown in \cref{fig:FM_DSF}(a,b) for $\mathbf{k} = \Gamma$ and K, respectively. Starting from the KSL at $h=0$, the applied field leads to broadening of the intense band of spin excitations that appear just above the two-flux gap in the range 
 $\omega \sim  0.1-0.3 \ |K|$. This broadening can be attributed to the field-induced dispersion for the fluxful excitations, which ultimately leads to the closure of the spin gap at $h_\text{c}^\text{FM} \sim 0.02|K|$. This signals a strong mixing between different flux sectors, and a breakdown of the zero-field topological order. 
  Coming from high-field, semiclassical spin-wave approaches would suggest a softening of magnons at all wave vectors on approaching the spin liquid\cite{mcclarty2018topological}. As a result, the spin excitation gap may close 
everywhere in $\mathbf{k}$-space simultaneously. For $h>h_\text{c}^\text{FM}$, the slope of the magnon energies with respect to field is smallest at $\mathbf{k}=\Gamma$ (compare \cref{fig:FM_DSF}(a) and (b)), as the minimal energy magnons exist at the zone center in the limit of large field.

\begin{figure}
\includegraphics[width=\linewidth]{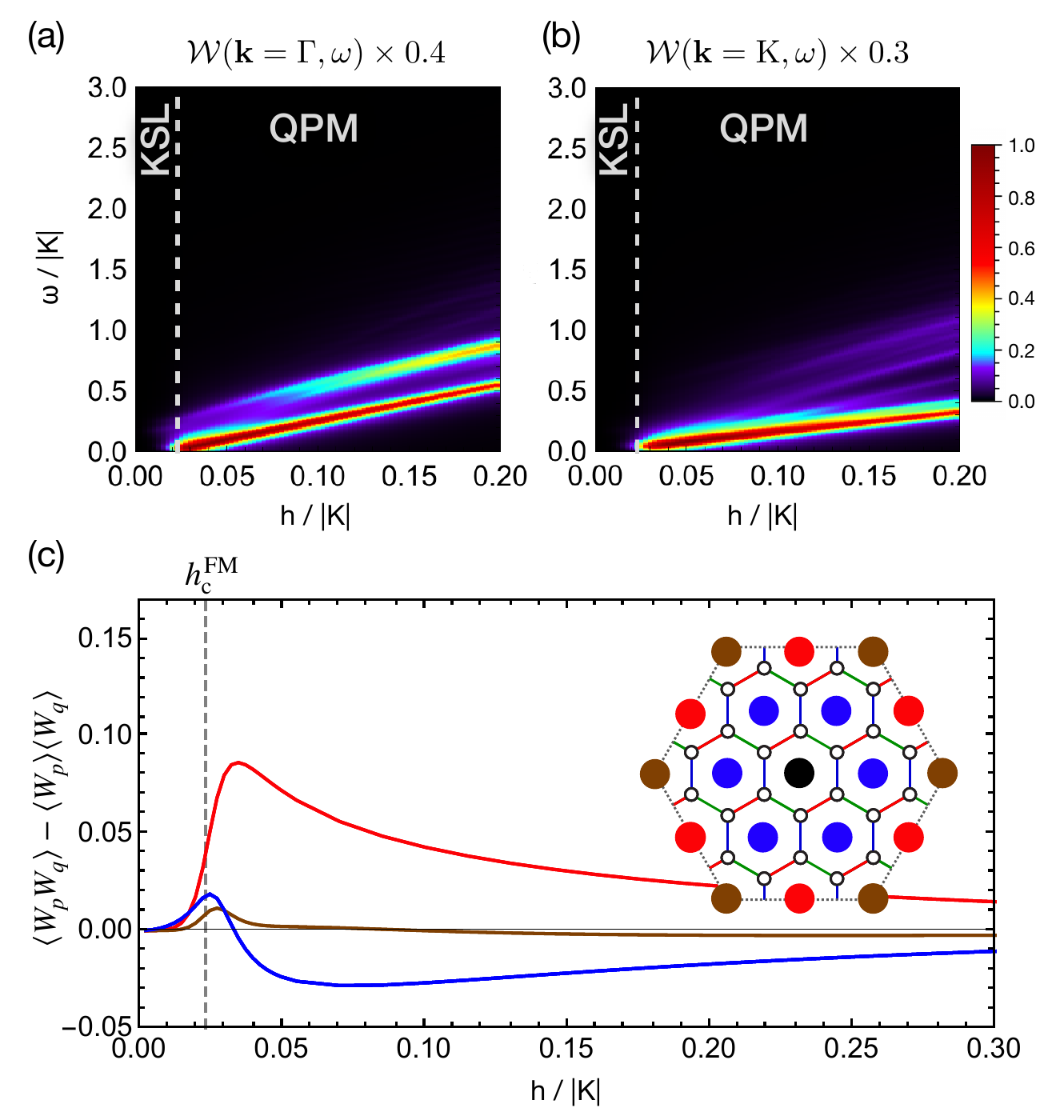}
\caption{
Flux correlations in the FM Kitaev model under $[111]$-field.  (a,b)~Dynamical flux-structure factor $\mathcal W(\mathbf k,\omega)$. 
 (c)~$\braket{W_pW_q}-\langle{W_p}\rangle\langle{W_q}\rangle$, where each line color refers to the relative position of the respective plaquette $p$ to the black plaquette $q$ shown in the inset.  
 }
\label{fig:FM_WP}
\end{figure}

The emergence of fluxes under field can be seen in the evolution of the average
flux density $\braket{n_p}$ shown in \cref{fig:FM_DSF}(c). While fluxes remain
to be nearly absent in the KSL until the critical field $h_\text{c}^\text{FM}$,
the average flux density and local flux density fluctuations measured by
 $\sigma_p=\sqrt{\braket{{n_p}^2}-\braket{n_p}^2}$ rapidly increase at the transition into the QPM. Remarkably,
$\braket{n_p}$ stays significantly below the limit\footnote{For a classical
product state with collinear spins (such as the fully polarized state), the
flux density is constrained to be close to 0.5, with $\frac{13}{27} \leq
\braket{n_p}\leq \frac{14}{27}$. Non-collinear classical states are less
constrained, and may have $\frac{1}{3} \leq \braket{n_p}\leq \frac{2}{3}$.} for
the classical polarized state, $\lim_{h\to \infty} \braket{n_p} = \frac{13}{27}
\simeq 0.48$, for a wide range of field strengths, implying significant quantum
fluctuations in the QPM. The latter are also evident from long-range spin-spin
correlations developing only gradually above $h_\text{c}^\text{FM}$,
cf.~\cref{fig:FM_DSF}(d).    

Since $W_p$ commutes with $H_K$, the dynamics of
the fluxes are mostly controlled by the field strength. In the asymptotically polarized region
at high-field, the dynamical flux-structure factor
$\mathcal{W}(\mathbf{k},\omega)$ (see \cref{fig:FM_WP}(a,b) 
for $\mathbf{k} = \Gamma, \text{K}$ respectively) features a
series of excitation bands corresponding to $n$-spin flips that do no alter
$\langle S_i^\gamma S_j^\gamma\rangle$. On approaching $h_\text{c}^\text{FM}$ from above, these excitations collapse
into a narrow frequency range on the scale of the zero-field flux gap
$\Delta_\text{f}$. Importantly, since $h_\text{c}^\text{FM},\Delta_\text{f} \ll
|K|$, the flux dynamics near the critical field are exceedingly slow compared
to the time scales for other excitations. As a result, the finite flux-density 
state just above the critical field may be discussed in terms of nearly
static fluxes, making it useful to consider the real-space static flux-flux correlations.

The static flux-flux correlations $\langle W_p W_q\rangle - \langle W_p \rangle \langle W_q\rangle$ are shown in \cref{fig:FM_WP}(c). Deep in the KSL, the magnetic field creates virtual pairs of fluxes on neighboring plaquettes, which then may hop to adjacent plaquettes. However, flux pairs remain confined due to the finite flux gap. As a result, fluctuations around the flux-free ground state lead to positive real-space correlations that decay with increasing plaquette separation. In contrast, for $h>h_\text{c}^\text{FM}$, the correlations are markedly different, likely reflecting effective interactions between fluxes that exist in finite density. The energetics of different flux configurations was studied first by Kitaev\cite{kitaev2006anyons}, and later by Lahtinen {\it et~al.}\cite{lahtinen2011interacting,lahtinen2012topological,lahtinen2014perturbed}. In analogy with vortices in $p+\imag p$ superconductors, each flux binds a Majorana $c$-fermion\cite{kitaev2006anyons,theveniaut2017bound} under applied field. Minimizing the energy of the Majorana bound states for pairs of fluxes leads to effective flux-flux interactions, which prefer that two fluxes are located on second neighbor plaquettes parallel to a bond, for example. The correlations observed near the critical field are consistent with flux patterns that minimize these interactions. Indeed, provided that the dynamics of the fluxes remain slow compared to the $c$-fermions, the essential effects of such bound states are likely to be preserved. While the spatial range of flux-flux correlations is difficult to diagnose from finite-size calculations, we note that a state with true long-range flux order would necessarily break additional lattice symmetries, and would therefore be distinct from the fully polarized state. Since 
no signatures of an additional phase transition have been detected\cite{zhu2018robust,gohlke2018dynamical}, it is likely that the flux-flux correlations retain a finite range in the thermodynamic limit. Therefore, the large flux-flux correlations observed in these  calculations do not appear to reflect the formation of a long-range ordered flux (vison)
 crystal of the type studied in Ref.\ \onlinecite{zhang2019vison}.

 The fate of the matter fermions can be diagnosed from dynamical bond correlations. We therefore discuss the spin-conserving channel of the resonant inelastic x-ray scattering (RIXS) intensity $\RIXS(\mathbf k,\omega)$, as defined in \cref{eq:RIXSintensity}. For completeness, we note that the form of the operator in Eq. (5) neglects additional contributions to the response that can appear at finite field related to differences in $g$-tensors between core and valence shell states. For a direct RIXS process in the first-order fast-collision approximation (employed in Ref.~\onlinecite{halasz2016resonant}), the RIXS operator gains an additional single-spin term under field, with relative magnitude 
 $\sim \left|(\mathbf g_\text{c}-\mathbf g_\text{v})\cdot \mathbf h\right|/|K|$, where $\mathbf g_\text{c}$ and $\mathbf g_\text{v}$ are the $g$-tensors of the core and valence shells
 that are involved in the RIXS process. Under the assumption that these $g$ tensors are similar, and for fields $h$ smaller than $|K|$, the main contribution to the spin-conserving RIXS intensity should therefore still be described by $\RIXS(\mathbf k,\omega)$ under finite fields. 

\begin{figure}
\includegraphics[width=\linewidth]{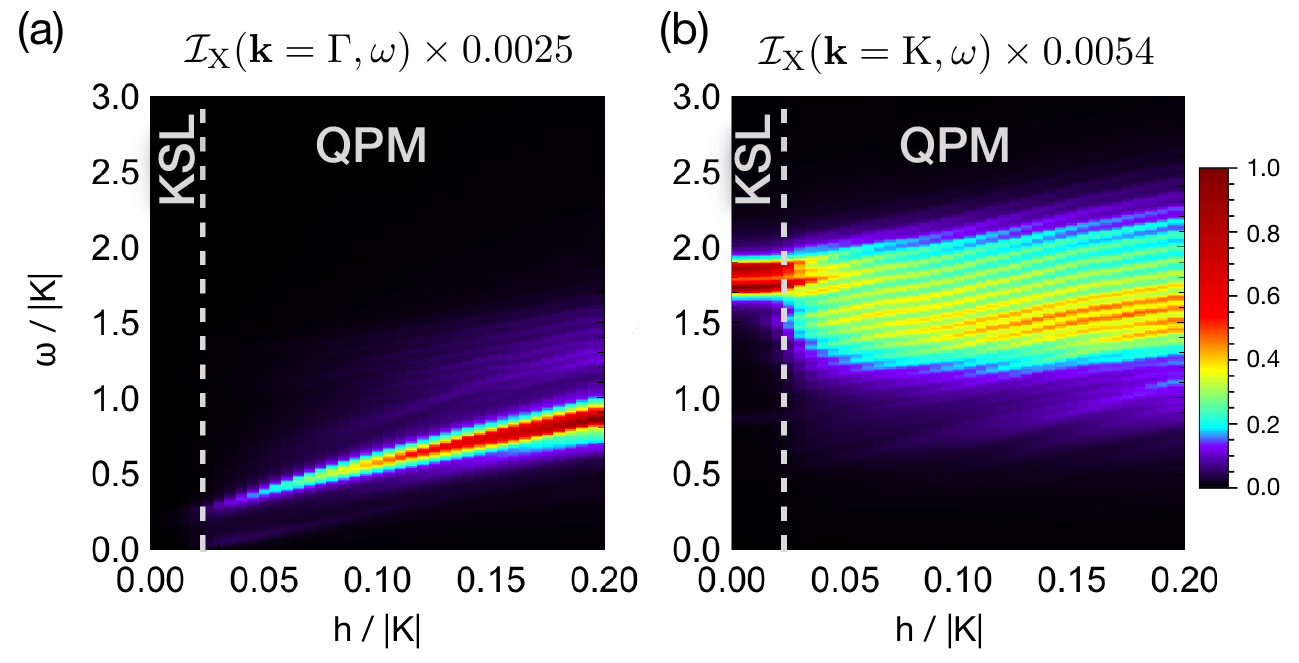}
\caption{
Spin-conserving RIXS intensity $\RIXS(\mathbf k, \omega)$ in the FM Kitaev model under $[111]$-field. 
 }
\label{fig:FM_RIXS}
\end{figure}

 Exact diagonalization results for $\RIXS(\mathbf k,\omega)$ are shown in \cref{fig:FM_RIXS} (a) and (b)
 for $\mathbf k=\Gamma$ and $\text K$ respectively. 
In the zero-field KSL, the majority of the spectral weight is represented by two-fermion excitations at energies $\omega \gtrsim 0.8 |K|$ at all wave vectors, as in \cref{fig:FM_RIXS}(b). The $\mathbf{k}$-points $\Gamma$, and $\Gamma'$ are an exception, as the intensity vanishes at zero field due to $[B_{\mathbf k=\Gamma,\Gamma'},H_K]=0$ (\cref{fig:FM_RIXS}(a)). The discreteness of these features in our calculations can be attributed to finite size effects, but their energy range agrees with exact calculations at zero field\cite{halasz2016resonant}. Under finite field, the sharp excitations of the KSL significantly broaden, consistent with strong scattering from the finite density of fluxes introduced by $h$. 
 As a result, the matter fermions cease to be well-defined excitations almost immediately upon entering the QPM phase above $h_\text{c}^\text{FM}$. In contrast to the AFM Kitaev model discussed below, the broad excitation bands, that emerge above $h_\text{c}^\text{FM}$, begin at frequencies similar to their corresponding zero-field two-fermion excitations. 
At high fields strengths $h \gg h_\text{c}^\text{FM}$ , the QPM can eventually be described as asymptotically polarized. Since
 the bond-bond correlations probe multi-spin flip excitations, they become increasingly expensive under applied field and,
 accordingly, the broad bands of excitations shift to higher energies with increasing $h$.

\subsection{Antiferromagnetic Kitaev model\label{sec:AFM_uniform}}

\begin{figure}
\includegraphics[width=\linewidth]{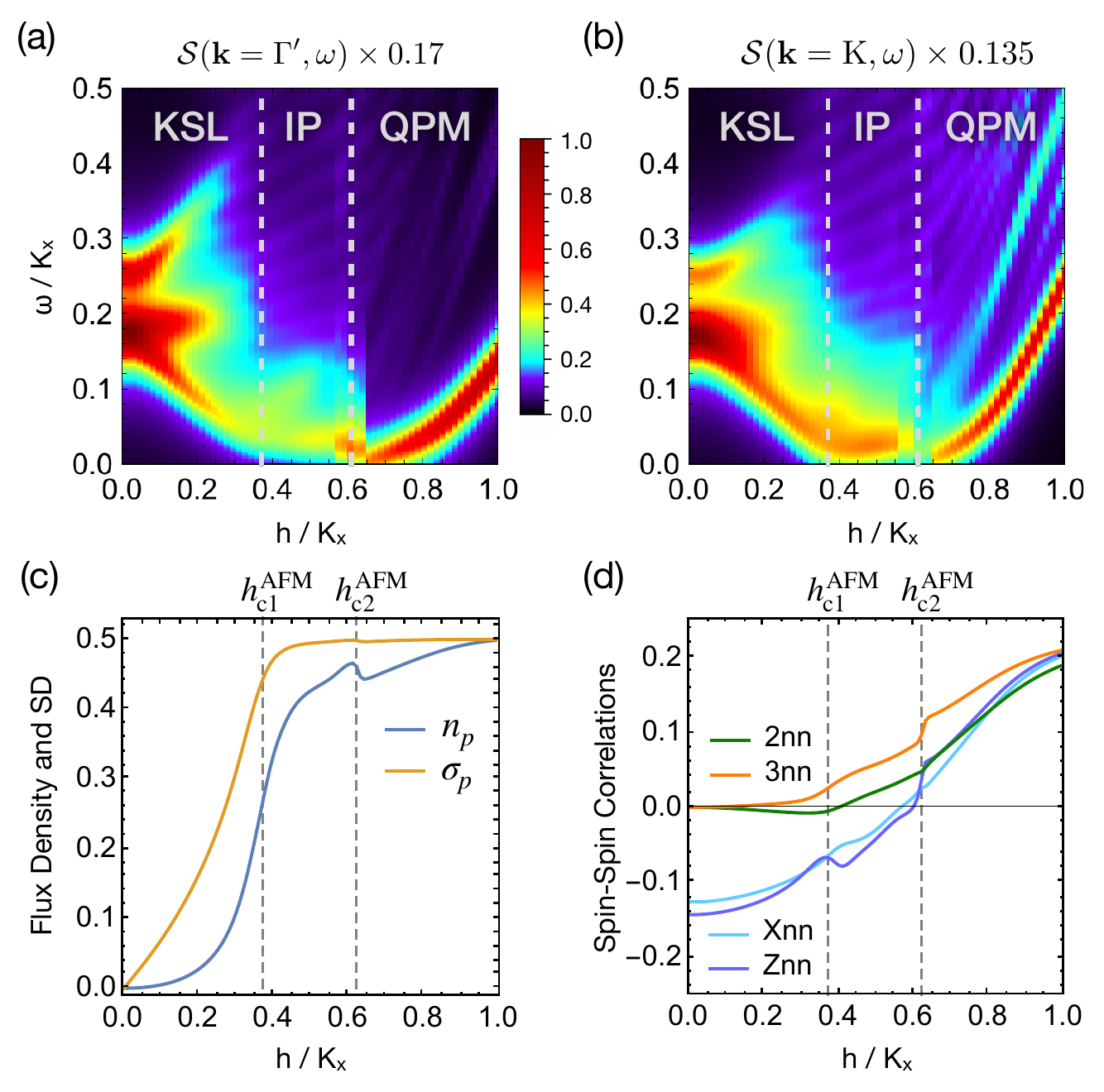}
\caption{
AFM Kitaev model with $K_x=K_y, K_z=1.05K_x$  under $[111]$-field.  (a,b)~Low-frequency dynamical spin-structure factor. 
 (c)~Flux density $\braket{n_p}$  and standard deviation (SD) of the local flux density $\sigma_p=\sqrt{\braket{{n_p}^2}-\braket{n_p}^2}$. (d)~Static spin-spin correlations in real space. Xnn and Znn denote $\braket{\mathbf S_i \cdot \mathbf S_j}$ on X- and Z-bonds, respectively.
 }
\label{fig:AFM_DSF}
\end{figure}

\begin{figure}
\includegraphics[width=\linewidth]{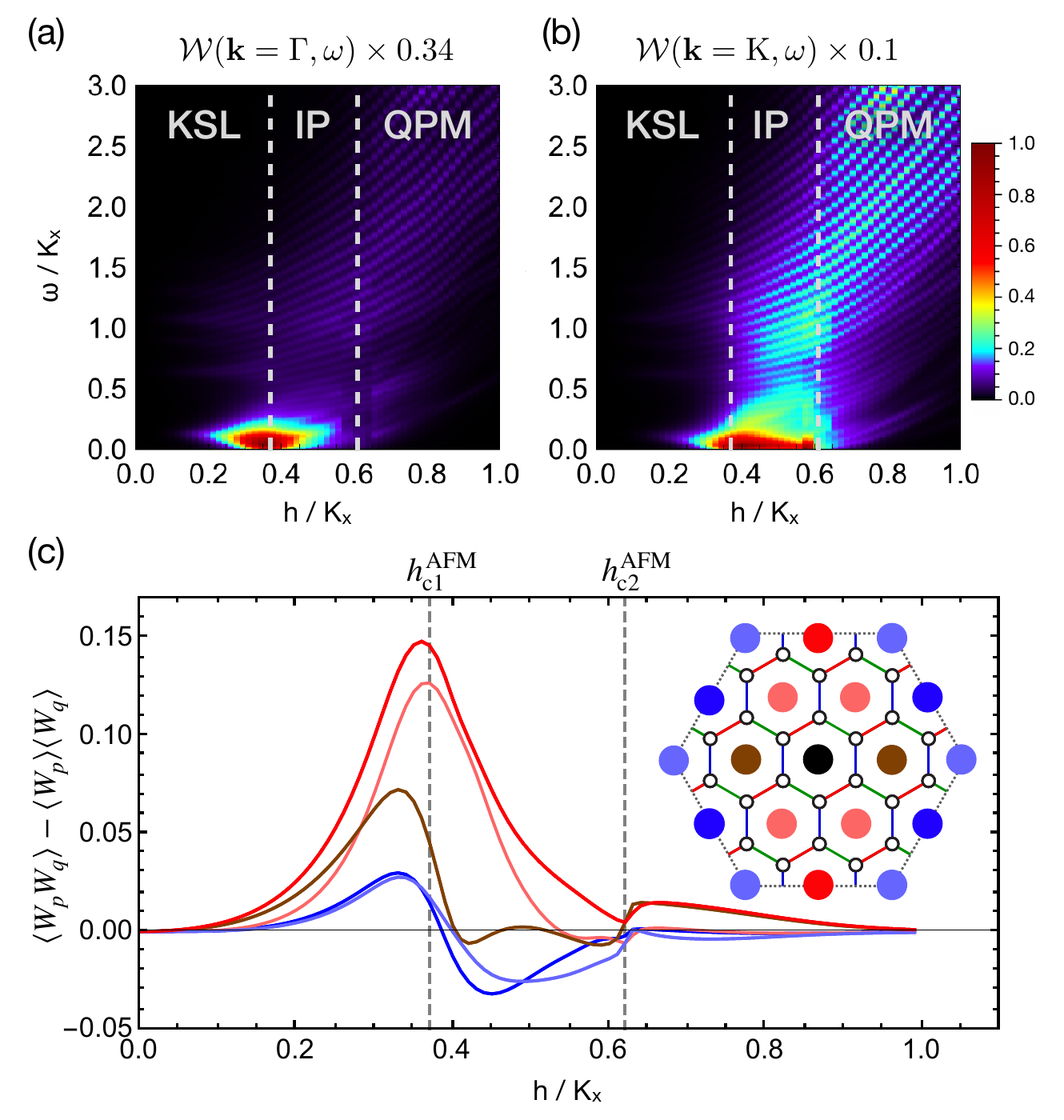}
\caption{
Flux correlations in the AFM Kitaev model with $K_x=K_y, K_z=1.05K_x$  under $[111]$-field. 
 }
\label{fig:AFM_WP}
\end{figure}

\begin{figure}
\includegraphics[width=\linewidth]{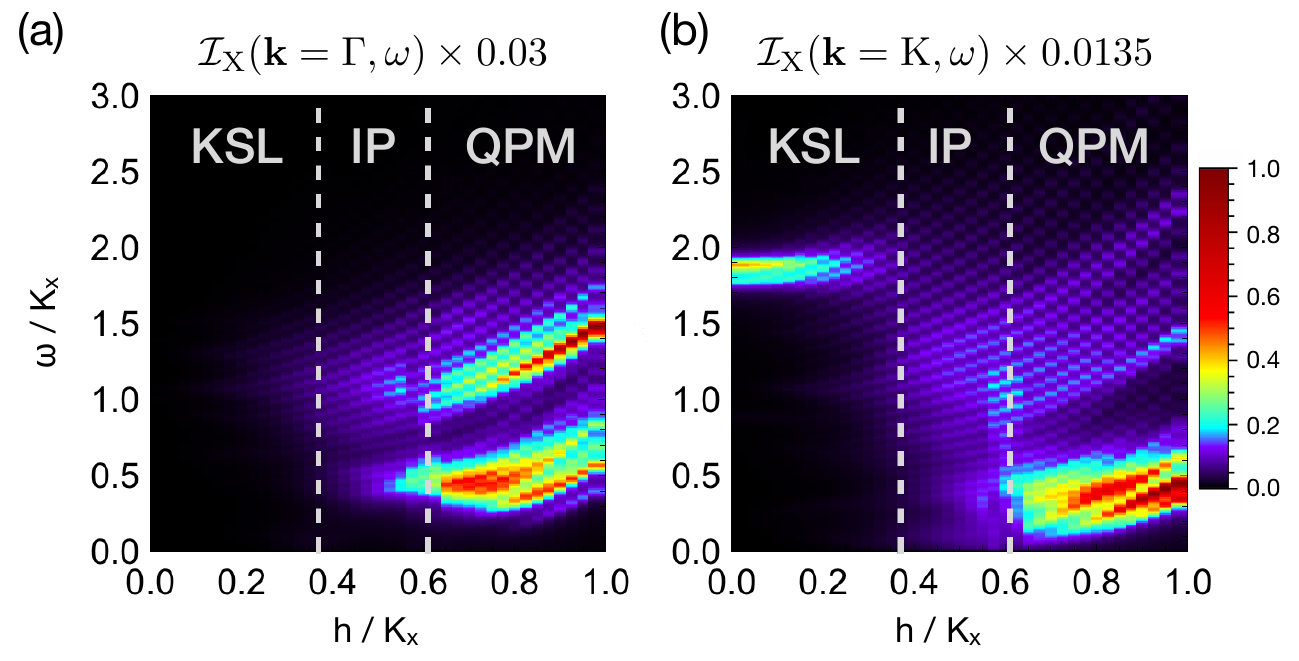}
\caption{
Spin-conserving RIXS intensity $\RIXS(\mathbf k, \omega)$ in the AFM Kitaev model with $K_x=K_y, K_z=1.05K_x$ under $[111]$-field. 
 }
\label{fig:AFM_RIXS}
\end{figure}

We now consider the AFM Kitaev model in uniform [111] field. To mitigate some finite-size effects in the {\it intermediate
phase}\footnote{\label{footnote:artifacfts} 
On the high-symmetry 24-site cluster under uniform [111]-fields, $\partial^2_h E_0$ and $\chi_h$  show multiple anomalies within the IP region, reflecting level crossings between nearly degenerate discrete states. However, these level crossing occur without qualitative
changes in any static or dynamic observable. Previous studies have mitigated such effects by (i)~employing clusters that do not respect $C_3$ symmetry\cite{zhu2018robust,gohlke2018dynamical,jiang2018field,zou2018field,patel2018magnetic,jiang2019field}, (ii)~rotating $\mathbf h$ slightly away from $[111]$\cite{hickey2019emergence}, or (iii)~introducing different coupling strengths on different bond types.}, 
 we slightly break $C_3$ symmetry by choosing the coupling strength on Z-bonds ($K_z$) stronger than that on X- and Y-bonds: $K=K_x=K_y$, $K_z=1.05K_x$. Dynamic and static response functions computed via ED are shown in
 \cref{fig:AFM_DSF,fig:AFM_WP,fig:AFM_RIXS}.

For the dynamical spin-structure factor shown in \cref{fig:AFM_DSF}(a,b)
at $\mathbf{k} = \Gamma^\prime, \text K$, we reproduce the results of Ref.~\onlinecite{hickey2019emergence}. At small $h$, the intense excitations at the flux gap broaden significantly, with the lower bound reaching nearly zero frequency at $h_{\text{c}1}^{\rm AFM} \approx 0.4K$. Unlike a transition to a spin-ordered phase \cite{gotfryd2017phase}, the spin gap appears to close everywhere in $k$-space simultaneously at $h_{\text{c}1}^{\rm AFM}$, so that clear magnetic Goldstone modes do not emerge at low frequencies. The spin gap remains small within ED resolution up to the second critical field $h_{\text{c}2}^{\rm AFM} \approx 0.6K$. This observation has previously been interpreted as the presence of gapless spin excitations in the IP in the thermodynamic limit\cite{hickey2019emergence,gohlke2018dynamical}.

Similar to the FM model, the vanishing energy difference between different flux
sectors at $h_{\text{c}1}^{\rm AFM}$ allows fluxes to proliferate. As a result,
the first critical field marks a rapid increase in the average flux density
$ n_p$, as well as in local flux-density fluctuations 
$\sigma_p$, see \cref{fig:AFM_DSF}(c).
In contrast, both quantities change very little at the second critical field
$h_{c2}^\text{AFM}$.

In the IP, the dynamical flux-structure factor $\mathcal W(\mathbf k, \omega)$, shown in \cref{fig:AFM_WP}(a,b), displays a broad continuum  in the energy range $\omega \sim 0$ to $3K$. However, much of the spectral weight is concentrated at small frequencies $0 \lesssim \omega \lesssim 0.1 K$, on the same scale as the zero-field flux gap. 
This indicates that the flux fluctuations---while large in amplitude $\sigma_p$---occur primarily on relatively \textit{slow} time scales throughout the IP. 
This is in contrast to neighboring ordered phases of the KSL, which can be induced by e.g.\ an additional Heisenberg term, where we find the spectral weight of $\mathcal{W}(\mathbf{k},\omega)$ to be concentrated at higher frequencies, with negligible weight near $\omega \approx 0$. 
Likewise, in the QPM phase in \cref{fig:AFM_WP}(a,b) for $h>h_{\text{c}2}^{\rm AFM}$, the low-frequency intensity in $\mathcal{W}(\mathbf{k},\omega)$ is rapidly suppressed. Here, the opening of a gap at $h_{\text{c}2}^{\rm AFM}$ shifts all excitations to higher energies with increasing field. Similar to the FM model near $h_\text{c}^\text{FM}$, the IP features pronounced modulation of the static flux-flux correlations $\langle {W}_{p}{W}_{q}\rangle - \langle  W_p\rangle\langle  W_{q}\rangle$ in real space, shown in \cref{fig:AFM_WP}(c). In this case, the correlations show a stripy pattern, the orientation of which is selected by the choice of $K_z > K_x,K_y$. These correlations are a property of the IP, and are largely suppressed upon approaching $h_{c2}^\text{AFM}$.

The spin-conserving RIXS intensity $\mathcal{I}_\text{X} (\mathbf{k},\omega)$ of the AFM Kitaev model under field is  shown in \cref{fig:AFM_RIXS}.  Here, the response under field differs significantly from the FM
Kitaev model due to the negative sign of static bond correlations $\braket{S_i^\gamma S_j^\gamma}$ in the low-field KSL, cf.~\cref{fig:FM_DSF}(d) and \cref{fig:AFM_DSF}(d).
 Upon entering the IP at $h_{\text{c}1}^{\rm AFM}$, the discrete two-fermion excitations of the KSL observed in ED dissolve into a broad band with no distinct frequency or momentum dependence (compare Fig. \cref{fig:AFM_RIXS} (a) and (b)), confirming that the $c$-fermions are strongly perturbed by the presence of the fluxes. At $h \sim h_{\text{c}1}^{\rm AFM}$, a significant portion of spectral weight in $\mathcal{I}_{\text X} (\mathbf{k},\omega)$ remains at high frequencies. However, in contrast to the FM model, part of the broad band shifts downward with increasing $h$, as the magnetic field reduces the energy cost for flipping the signs of $\langle S_i^\gamma S_j^\gamma \rangle$. Finally, at $h = h_{\text{c}2}^{\rm AFM}$, the band is driven to $\omega \approx 0$ at all wave vectors.  Consistently, the static nearest-neighbor spin-spin correlations rapidly reverse sign upon leaving the IP, as shown in \cref{fig:AFM_DSF}(d).

\section{Stability of the Intermediate Phase in Extended Models\label{sec:general_fields}}
 \subsection{Non-Collinear Fields \label{sec:general_fieldsA}}
\begin{figure}
 \includegraphics[width=\linewidth]{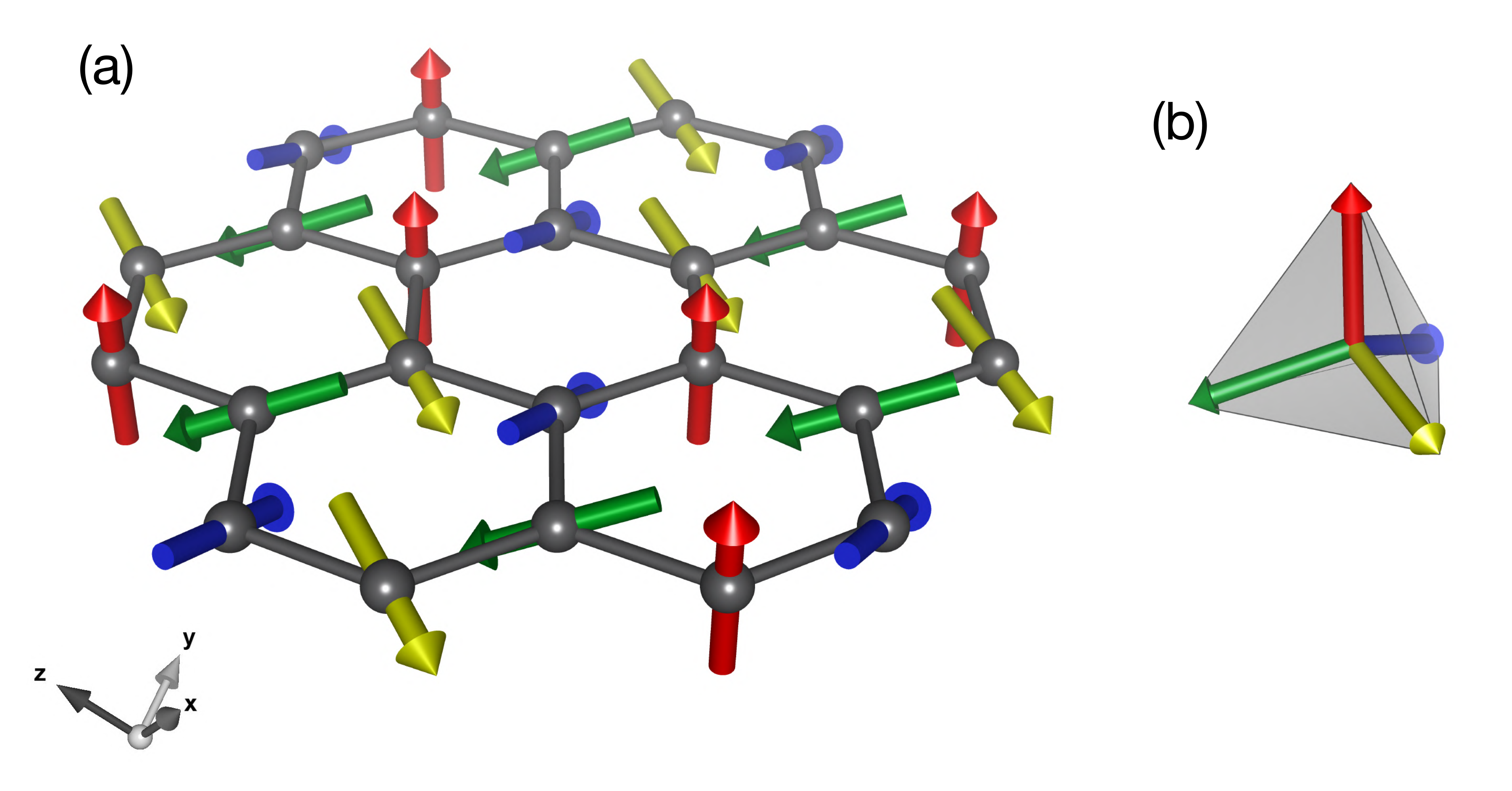}
\caption{(a) Site-dependent directions of the four-sublattice  tetrahedral
 field that acts on $H_K$ in the same way as a uniform field does on $-H_K$.  
The relative orientations of the local fields match those of the vertex corners of a tetrahedron, as shown in (b). 
}
\label{fig:tetrahedral_3d}
\end{figure}

 As discussed in Section \ref{sec:Kitaev_uni_field}, the presence of the IP in the the AFM Kitaev model 
under a uniform field can be anticipated from two observations: (i) A finite field rapidly induces flux density fluctuations (suppressing the KSL) at $h \sim \Delta_\text{f}$, but (ii) does not immediately lift the extensive classical degeneracy. For this reason, a quantum spin liquid phase at intermediate fields can be anticipated. This argument can be extended to include general site-dependent fields defined by:
 \begin{align}
\fieldterm= -\sum_i h_i^x S_i^x + h_i^y S_i^y + h_i^z S_i^z
\end{align}
For field configurations that do not couple to any classically degenerate state, the product of the $\gamma$-component of the local fields on all $\gamma$-bonds must satisfy
\begin{equation}
   \text{sgn}(h_i^\gamma h_j^\gamma) = \text{sgn}(K),\label{eq:necessary_condition}
\end{equation} 
which serves as a \textit{necessary} condition for the stabilization of the IP. This suggests that the IP may also be induced in the FM model for certain non-collinear field configurations satisfying $h_i^\gamma h_j^\gamma > 0$. In fact, as discussed in Appendix \ref{sec:symmetries}, the AFM Kitaev model under uniform [111] field is exactly dual to the FM model under the four-sublattice ``tetrahedral'' field pictured in Fig.~\ref{fig:tetrahedral_3d}. Similarly, the AFM model under the four-sublattice field is dual to the FM model under uniform [111] field.

These dualities allow us to consider a series of models that interpolate between those possessing the IP, and those where a direct transition occurs between the KSL and the QPM.
 To this end, we consider the phase diagram of the AFM Kitaev model with site-dependent fields:
\begin{equation}
  H(h,\phi) = H_{K,\text{AFM}} - h \sum_i \mathbf{h}_i(\phi)\cdot \mathbf S_i
  ,
  \label{eq:Htetrahedral_interpolation}
\end{equation}
with local field directions $\mathbf h_i(\phi)$ defined by the sublattice pattern in \cref{fig:transforms1}(d), with:
\begin{align}\label{eq:tetrafields_parametrization1}
   \mathbf h_i(\phi) &
    = 
    \frac { 1} { \sqrt { 3 } } 
    \left\{ \begin{array} { l l }
        { ( 1,1,1 ), } & { \text{if }i \in \text {sublattice  A} }
     \\ { ( v , u , v ), } & { \text{if }i\in \text {sublattice  B} } 
     \\ { ( v , v , u ), } & { \text{if }i\in \text {sublattice  C} }
     \\ { ( u , v , v ), } & { \text{if } i \in \text {sublattice  D} } 
  \end{array} \right.\\
   u(\phi) & = \cos ( \phi ) + \sqrt { 2 } \sin ( \phi ) , \label{eq:tetrafields_parametrization2}
 \\ v(\phi) & = \cos ( \phi ) - \sin ( \phi ) / \sqrt { 2 } . \label{eq:tetrafields_parametrization3}
\end{align}
For this choice, $\mathbf{h}_i(\phi)$ rotates directly between a uniform [111] field (for $\phi=0$), and the tetrahedral field shown in \cref{fig:tetrahedral_3d} (for $\phi=\arccos(-\frac13)\equiv \phiFM \simeq 0.61\pi$).  As introduced above, the model is dual to the FM Kitaev model in uniform [111] field for $\phi=\phiFM$. 

\begin{figure}
\includegraphics[width=\linewidth]{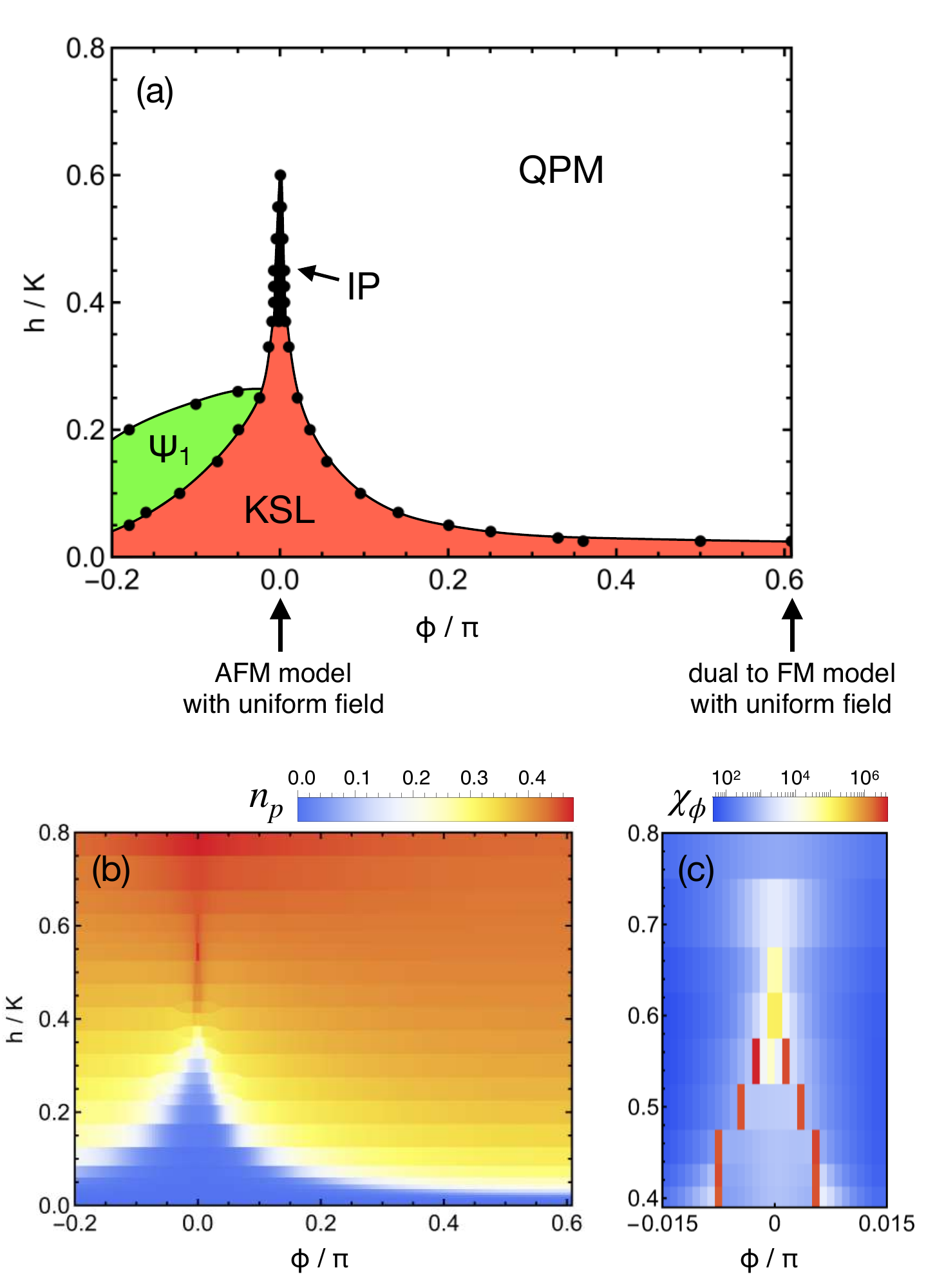}
\caption{
(a)~Phase diagram interpolating between the AFM Kitaev model under a
uniform $[111]$-field ($\phi=0$) and 
a model dual to the FM model under a uniform $[111]$-field ($\phi=\phiFM\simeq0.61\pi$) via \cref{eq:Htetrahedral_interpolation,eq:tetrafields_parametrization1,eq:tetrafields_parametrization2,eq:tetrafields_parametrization3}. Black points depict maxima in the fidelity susceptibility. (b)~$\mathbb Z_2$~Flux density (\cref{eq:flux_density}). (c)~Fidelity susceptibility (\cref{eq:fidelitysuscept}) with respect to $\phi$, close to the limit of uniform fields $\phi=0$. Calculated with $\delta \phi = 0.005\pi$. Note that the color scale is logarithmic. 
}
\label{fig:tetraB_phasediag}
\end{figure}

The phase diagram as a function of $\phi$ and $h$ obtained via ED is shown in \cref{fig:tetraB_phasediag}(a). Phase boundaries were identified from maxima in the second derivative of the ground state energy $-\partial^2_g E_0$ and in the fidelity susceptibility
\begin{equation}
  \chi_g = \frac{-2 \log\left(\left| \langle\psi(g) | \psi(g+\delta g )\rangle \right|\right) }{(\delta g)^2}
  \label{eq:fidelitysuscept}
  ,
\end{equation}
where we have varied both $g=h$ or $g=\phi$. The locations of these maxima are shown as black points in \cref{fig:tetraB_phasediag}(a).

We first focus on the region for $\phi\geq 0$. As shown in the phase diagram, we find that the IP is very unstable against tetrahedral rotations, surviving only for the narrow range $\phi\lesssim 0.005\pi =0.9^\circ$ in our $N=24$ calculations. This is well before the necessary condition \eqref{eq:necessary_condition} for the IP stops being fulfilled at $\frac\phiFM2\simeq 55^\circ$. For all angles $\phi > 0.9^\circ$, a direct transition between the KSL and QPM is observed, with critical field $h_\text{c}$ decreasing monotonically with rotation angle. The extent of the KSL and IP can be deduced from the behavior of the flux density, plotted in \cref{fig:tetraB_phasediag}(b). In the KSL, $n_p$ is suppressed, while in the IP this quantity is enhanced compared to adjacent phases. In \cref{fig:tetraB_phasediag}(c), we show the behavior of the fidelity susceptibility $\chi_\phi$ in the narrow region around $\phi \sim 0$. For $0.4\lesssim \frac{h}{K} \lesssim 0.6$, the transition between the IP and the polarized phase occurs via level crossings on sweeping $\phi$, signified by two divergencies in $\chi_\phi$ at very small negative and positive $\phi$. For $ h\gtrsim 0.6K$, the divergences meet and are rapidly suppressed.

While the above results suggest the IP has a very small---but finite---extent with respect to $\phi$, we note that the energy spectra in finite-size calculations are necessarily discrete. 
The level crossings shown in \cref{fig:tetraB_phasediag}(c) can therefore not happen instantly at $\phi=0$ in our calculations, but must appear after a finitely large perturbation to the Hamiltonian. Provided that the IP is gapless in the thermodynamic limit, as concluded in Refs.~\onlinecite{zhu2018robust,gohlke2018dynamical,hickey2019emergence,jiang2018field,zou2018field,patel2018magnetic}, the critical $\phi$~values may scale to zero as the finite-size gap closes. In our calculations on 24 sites, the narrow width of the IP is controlled by the relative scale of the energy gaps $\sim 0.01K$ between the ground state and lowest excited states at $\phi=0$. 
For this reason, we cannot rule out a scenario in which the IP is reduced to a line of critical points in the $(h,\phi)$ plane of \cref{fig:tetraB_phasediag}(a) in the thermodynamic limit, with no finite extent in the $\phi$-direction. This suggests an intriguing instability with respect to tetrahedral fields. These non-uniform fields carry momenta at wave vectors $\mathbf k=\text{X}$. If the IP is a $\mathrm U(1)$ spin liquid with spinon Fermi pockets around certain momentum vectors, as proposed in Refs.~\onlinecite{jiang2018field,zou2018field,patel2018magnetic}, 
   then the perturbing tetrahedral fields might couple states at Fermi pockets that are offset by nesting vectors $\mathbf q = \text X$, thereupon immediately opening a gap. This suggestion appears consistent with recent DMRG studies, which show a peak in the static spin structure factor at the X-points for the AFM model in [111] fields \cite{patel2018magnetic}. We note that such a strong nesting would lead to Peierls-like structural instabilities if spin-lattice couplings were considered \cite{hermanns2015spin}.

Finally, we comment briefly on a phase detected for negative $\phi$. The region of this additional phase (named $\Psi_1$ in \cref{fig:tetraB_phasediag}(a)) can be outlined clearly by maxima in $\chi_g$ and $-\partial_g^2 E_0$.  From the static spin-structure factor, we can not identify a dominant ordering wave vector. 
Within the discrete spectra of the ED calculation, the gap throughout the region of $ \Psi_1$ is on the order of $0.02K$, so that one could speculate about a potential gaplessness in the thermodynamic limit. 
In this respect, it has similarities to the IP, however we do not find $\Psi_1$ to be smoothly connected to the IP in parameter space. Furthermore, the flux density in $\Psi_1$ is significantly lower than in the IP, yet still distinctly above that of the KSL, see \cref{fig:tetraB_phasediag}(b).

\subsection{Extended Interactions
\label{sec:extendedinteractions}} 

To date, material realizations of
Kitaev-like Hamiltonians have been sought in a number of spin-orbital coupled
transition metal compounds including $\alpha$-RuCl$_3$, Na$_2$IrO$_3$, and
various polymorphs of Li$_2$IrO$_3$. In such materials, the minimal nearest
neighbor couplings can be
parameterized\cite{rau2014generic,winter2016challenges}:  
	\begin{widetext}
	\begin{equation}
  	  	H_{JK\Gamma\Gamma'} =
	 \sum_{\langle ij \rangle_\gamma} J\ \mathbf{S}_i \cdot \mathbf{S}_j + K\ S_i^\gamma S_j^\gamma + \Gamma\ \left( S_i^\alpha S_j^\beta + S_i^\beta S_j^\alpha\right) + \Gamma' \ \left( S_i^\alpha S_j^\gamma + S_i^\gamma S_j^\alpha + S_i^\beta S_j^\gamma + S_i^\gamma S_j^\beta \right),
	 \label{eq:H_JKGGp}
	\end{equation}		
	\end{widetext}
	where $\gamma\in\{x,y,z\}$ corresponds to the type of bond connecting sites $i$ and $j$, and $(\alpha,\beta,\gamma)$ is always a permutation of $(x,y,z)$. While the precise determination of 
 magnetic Hamiltonians for the candidate materials poses an ongoing challenge, the Kitaev exchange has been shown to be of ferromagnetic type (FM, $K<0$), on grounds of the 
idealized microscopic 
mechanism\cite{jackeli2009mott}, \textit{ab-initio} studies\cite{katukuri2014kitaev,yamaji2014first,kim2016crystal,winter2016challenges,yadav2016kitaev,wang2017theoretical,winter2017models,yadav2018strain} and experimental analyses\cite{chaloupka2015hidden,chaloupka2016magnetic,do2017majorana,koitzsch2017low,yamauchi2018local,cookmeyer2018spin,das2019magnetic}. 
While we have discussed in \cref{sec:general_fieldsA} and \cref{sec:symmetries} that the 
field-induced IP can be realized in the FM Kitaev model, this requires non-collinear and staggered fields that are unlikely to be available in real experiments. At first glance, this puts into question the relevance of the physics of the IP to real materials. 

\begin{figure}
\includegraphics[width=\linewidth]{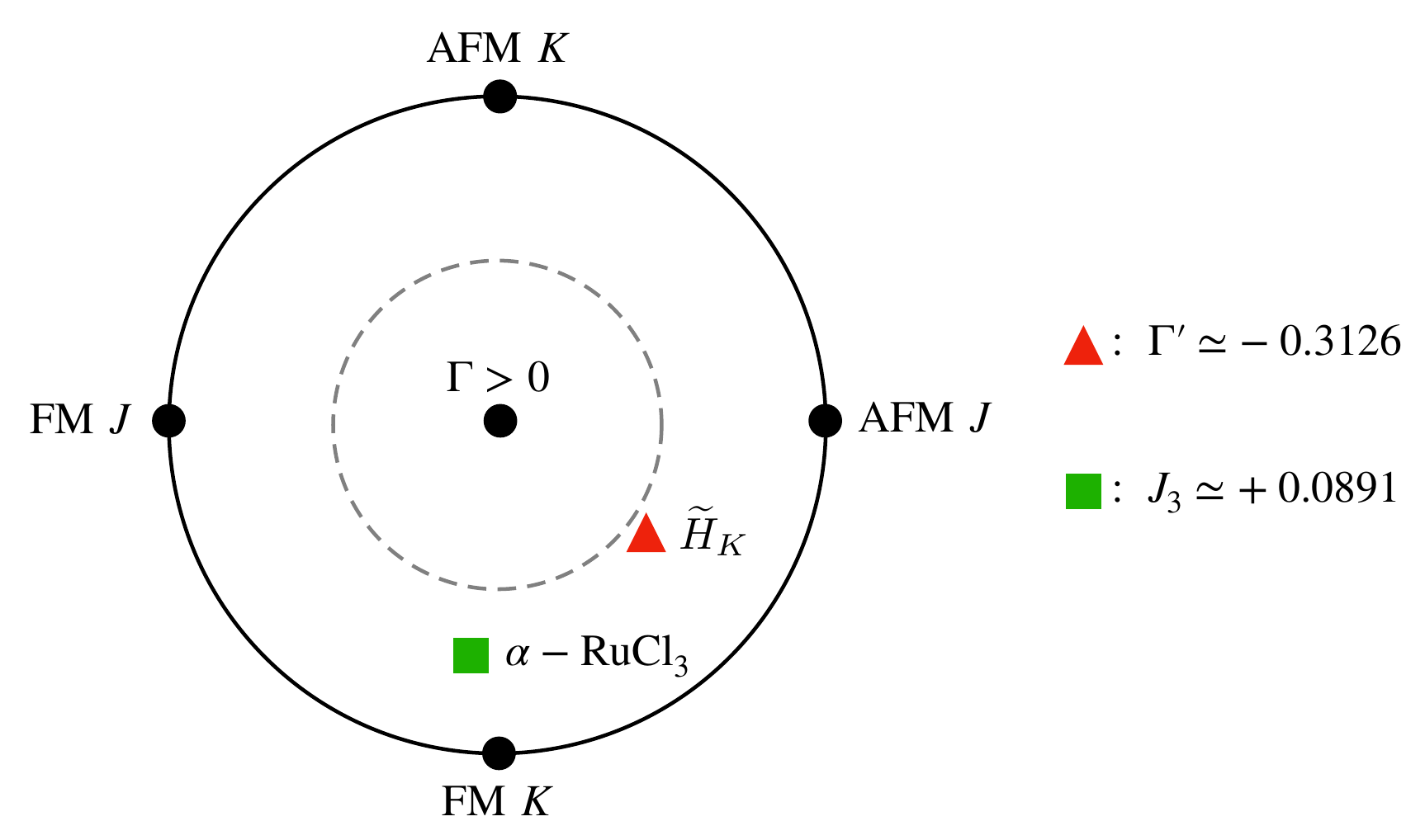}
\caption{
$(J,K,\Gamma)$-parameter space for $\Gamma>0$. We adopt the parametrization of Refs.~\onlinecite{rau2014generic,chaloupka2015hidden}, such that $J=\sin\vartheta \cos\varphi$, $K=\sin\vartheta\sin\varphi$, $\Gamma=\cos\vartheta$, where $\varphi$ is the polar angle and $\vartheta$ is proportional to the distance from the center, reaching $\vartheta=\pi/4$ at the dashed circle and $\vartheta=\pi/2$ at the outer circle. 
The black dots show models where only one type of coupling is present. 
The triangle and the square show the projected positions of the hidden AFM Kitaev model (\cref{eq:Hhidden}) and of the \rucl\ model (\cref{eq:Hrucl}), respectively. Additional couplings that are not 
encoded in $(\varphi,\vartheta)$
 are given on the right side in units of $\sqrt{J^2+K^2+\Gamma^2}$. 
}
\label{fig:jkg_space}
\end{figure}

From this viewpoint, it is useful to note the presence of {\it hidden} Kitaev
points in the extended  {$(J,K,\Gamma,\Gamma')$-parameter} space
(cf.~\cref{eq:H_JKGGp}), which are implied by transformations discussed in
Ref.~\onlinecite{chaloupka2015hidden}. In particular, a $\pi$-rotation of all
spin-operators around the [111] axis is defined by: \begin{equation}
  \mathcal R : (\tilde x, \tilde y, \tilde z)\transposed = 
\frac13  \begin{pmatrix}
-1x+2y+2z
\\ +2x-1y+2z
\\ +2x+2y-1z
\end{pmatrix}
.
\end{equation}
Applying this transformation
 to the pure AFM Kitaev model $H_K$ ($K=+1$) leads to a Hamiltonian $\widetilde H_K=\mathcal R H_K \mathcal R^{-1}$, 
that is of the ($J,K,\Gamma,\Gamma'$)-form 
 with parameters
\begin{equation}
  \widetilde H_K: \quad (\tilde{J},\tilde{K},\tilde{\Gamma},\tilde{\Gamma}') = \left(+\frac{4}{9},-\frac{1}{3},+\frac{4}{9},-\frac{2}{9}\right).
  \label{eq:Hhidden}
\end{equation}
Importantly, the signs of all anisotropic couplings (i.e. $K<0,$ $\Gamma > 0$ and $\Gamma'<0$) at this hidden AFM Kitaev point are compatible with microscopic mechanisms relevant to known Kitaev materials\cite{rau2014generic,winter2017models}. Such couplings are, in principle, realizable in real materials. The static $\mathbb Z_2$ gauge field for the KSL ground state of $\widetilde H_K$
 is related to new flux operators $\widetilde W_p=\mathcal R W_p \mathcal R^{-1}$. Since $\mathcal R$ commutes with the Zeeman term of a $[111]$ field,  $\widetilde H_K$
 hosts the IP under \textit{uniform} fields. 
In \cref{fig:jkg_space}, we show the position of the hidden AFM Kitaev point
$\widetilde H_K$ projected into the $(J,K,\Gamma)$-parameter space. 

Since both the KSL and the IP at the original AFM Kitaev point have finite extents when adding additional interactions to $H_K$ \cite{chaloupka2010kitaev,rau2014generic,hickey2019emergence} these phases must also have finite extent around the \textit{hidden} Kitaev model $\widetilde H_K$. It is therefore interesting to explore as to what extent these phases might be proximate to the phases of real materials. To that end, we take as a representative for real materials the \textit{ab-initio}-guided minimal model of Ref.~\onlinecite{winter2017breakdown} for $\alpha$-RuCl$_3$: 
\begin{equation}
\hrucl: \ (J, K, \Gamma, J_3 ) = (-0.5, -5, +2.5, +0.5) \text{ meV} 
   \label{eq:Hrucl}
  ,
\end{equation}
where $J_3$ stands for third-nearest-neighbor Heisenberg coupling. We then consider models:
\begin{equation}
H(\xi,\mathbf h) = (1-\xi)\, \widetilde  H_K + \xi \frac{1}{C}  \hrucl - \sum_i \mathbf h \cdot \mathbf S_i
\label{eq:hiddeninterpolation}
,
\end{equation}
which interpolate between the hidden AFM Kitaev point ($\xi=0$) and the \rucl\ model ($\xi=1$) under uniform fields $\mathbf h$.  
We set $C=8.5\text{ meV}$ for comparable energy scales. 
 \hrucl\ has been shown to be consistent with many experimental aspects of \rucl\ \cite{winter2017breakdown,ponomaryov2017unconventional,wolter2017field,winter2018probing,riedl2018saw,Note3}. 
\nocite{cookmeyer2018spin,lampen2018field,wu2018magnons} 
While the model can certainly still be further fine-tuned\footnote{\label{footnote:adjustedmodel}
The studies of 
Refs.~\onlinecite{cookmeyer2018spin,lampen2018field,wu2018magnons}  found further measurements well reproduced by the model with slightly adjusted parameters. 
}, we assert that the relative direction in parameter space between $\widetilde H_K$ and \rucl\ can be described reasonably well with \hrucl.  
 
 \begin{figure*}
  \includegraphics[width=1.00\linewidth]{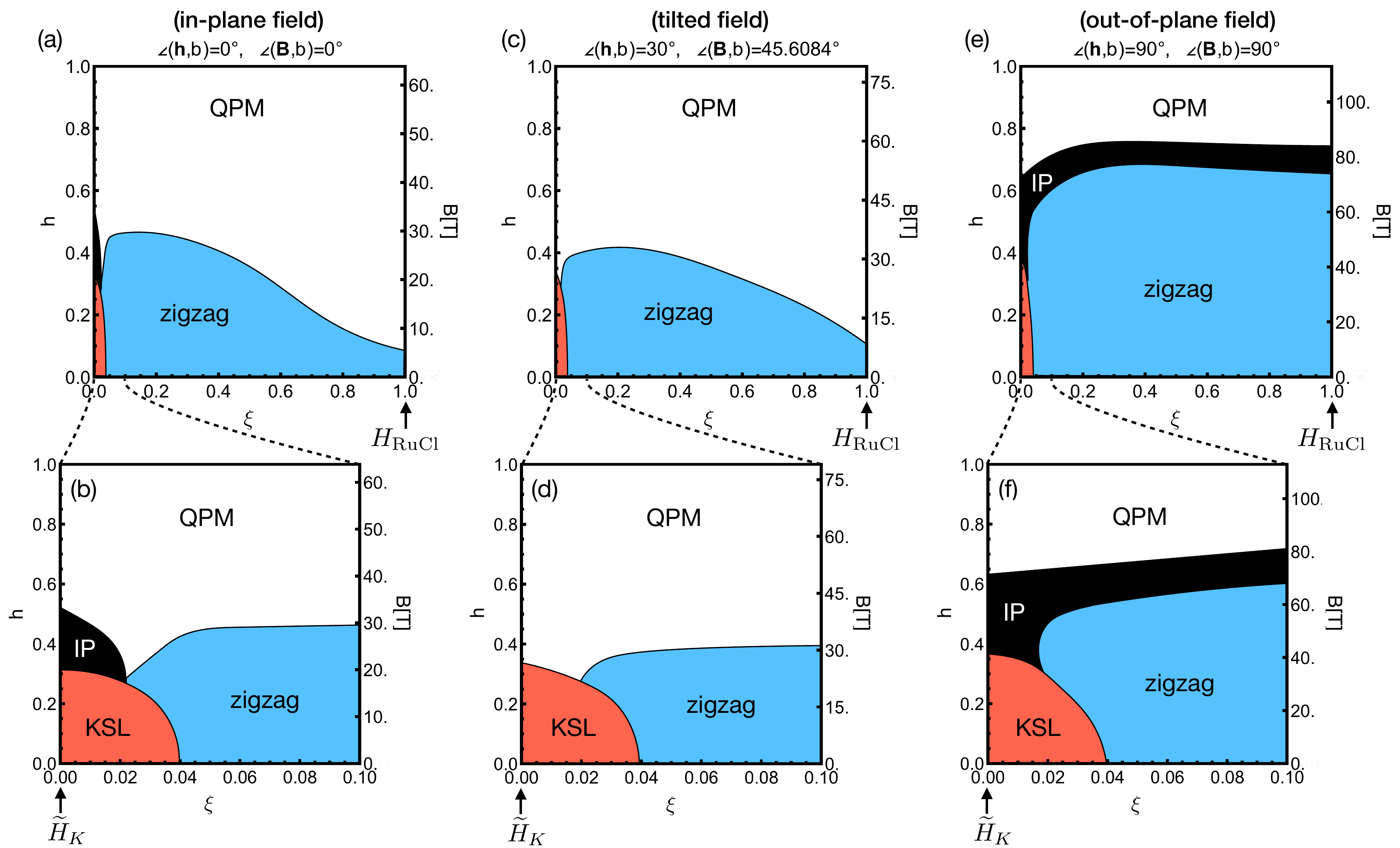}
  \caption{Phase diagrams of $H(\xi,\mathbf h)$ from \cref{eq:hiddeninterpolation} for uniform fields with directions between $b=[1\bar10]$ (in-plane) and $c^*=[111]$ (out-of-plane). $\xi=0$ corresponds to the hidden AFM Kitaev point $\widetilde H_K$ of \cref{eq:Hhidden}, and $\xi=1$ to the \rucl\ model 
   \hrucl. 
  The labels on the left axes show the energy scale of the Zeeman term as it enters \cref{eq:hiddeninterpolation}. The labels on the right axes show the corresponding external field strengths 
  $B =C \mu_\text{B}^{-1} |\mathbf g^{-1} \mathbf h|$ 
  in Tesla, when converting to the energy scale of \rucl\ and assuming the anisotropic $g$ tensor $\mathbf g$ of Ref.~\onlinecite{winter2018probing}. Phase boundaries are determined by peaks in $\partial^2_g E_0$ and in $\chi_g$. The quantum paramagnetic (QPM) phase is smoothly connected to the fully polarized state of the  $h\rightarrow \infty$ limit. (a,b)~Field along $[1\bar10]$, which corresponds to the crystallographic $b$ axis of \rucl. (c,d)~$\mathbf h$ tilted from $b$ towards $[111]=c^*$ by $30^\circ$. Assuming the mentioned $g$ tensor, this corresponds to an external $B$-field tilted by $\sim 45^\circ$. (e,f)~$\mathbf h \parallel [111] = c^*$.
  \label{fig:hiddenruclPD} 
  }
\end{figure*}

In \cref{fig:hiddenruclPD}~(a-f), we show the ground state phase diagrams of 
$H(\xi,\mathbf h)$ for three different directions of $\mathbf h$, going from in-plane ($\mathbf h \parallel b =[1\bar10]$) to out-of-plane ($\mathbf h \parallel c^* = [111]$) fields. Phase boundaries were determined from ED on the high-symmetry 24-site cluster, by tracking maxima in $\chi_g$ (\cref{eq:fidelitysuscept}) and in $-\partial^2_g E_0$ with $g=h$ or $g=\xi$. 
The physical $B$-field in Tesla is related to our natural units as $\mathbf B =C \mu_\text{B}^{-1} \mathbf g^{-1} \mathbf h$,
 shown on the right-hand axes 
 in \cref{fig:hiddenruclPD} for the $g$-tensor estimate of Ref.~\onlinecite{winter2018probing}.

On the described path through parameter space, the zigzag phase of $\hrucl$ is found to be a (direct) neighbor of the hidden KSL phase, meeting it at $\xi \simeq 0.04$ for $h=0$. 
With increasing field strength, this phase boundary bends towards smaller $\xi$, see Figs.~\ref{fig:hiddenruclPD}(b,d,f). Hence, no model exists on the path where an external field can suppress zigzag order in favor of a field-induced KSL state, unlike the path considered in Ref.~\onlinecite{gordon2019theory}, where the KSL can be field-induced for field directions close to $[111]$.  
For the models considered here, the stability of the KSL region is 
found to be qualitatively independent of field direction.

In contrast, the stability of the field-induced IP depends crucially on the field direction.  Even for $\xi = 0$ at the pure (hidden)
Kitaev model, the IP is not induced by uniform fields near to a cubic axis (e.g. [100]), as is the
case for the direction shown in \cref{fig:hiddenruclPD}(c,d). 
For most field directions where the IP is observed, it is found to be less stable than the KSL against additional interactions introduced by a finite $\xi$, as in \cref{fig:hiddenruclPD}(a,b),
 so that it cannot generally be field-induced from a zigzag ground state either. 
Remarkably, however, we find exclusively for the out-of-plane field-direction $[111]=c^*$, that the IP extends far through parameter space, reaching even the \hrucl\ model at $\xi=1$ at high fields, see \cref{fig:hiddenruclPD}(e,f). This result applies only for fields very near to the [111] direction\footnote{The magnetic torque response of \hrucl\ is found to be nearly unaffected by the presence of the IP on sweeping the field angle, and was thus not discussed in the study of Ref.~\onlinecite{riedl2018saw} on the same model}. Upon rotating $\mathbf h$ by small angles of $\lesssim 10^\circ$ away from $[111]$, the extended region shrinks quickly, such that the extent of the IP with respect to $\xi$ becomes qualitatively similar to that in \cref{fig:hiddenruclPD}(b).

 The peculiarity of the $[111]$ field direction for the stabilization of a field-induced 
phase in models with zigzag order hints at importance of maintaining $C_3$ symmetry. It may be worth noting, provided that $C_3$ symmetry is maintained, that a direct continuous transition between the zigzag and polarized phases is unlikely to occur in the thermodynamic limit. This is because the symmetry group of the broken symmetry zigzag ordered phase is not a maximal subgroup of the higher symmetry polarized state\cite{ascher1977symmetry,ascher1977permutation}. As a result, the field-induced transition is likely to be either first order (as observed for the classical model\cite{janssen2017magnetization} at $\xi=1$), or feature one or more intermediate phases (as observed for zigzag phases in the classical Kitaev-Heisenberg model \cite{janssen2016honeycomb}). In finite-size calculations on the quantum model\cite{hickey2019emergence,jiang2019field} however, the field-induced IP  appears to lack long-range magnetic 
order, and therefore is inconsistent with either scenario observed in the classical limit. 
Finally, we remark the presence of level crossings between quasi-degenerate states within the IP region on the high-symmetry cluster for $[111]$ fields\cite{Note2}. 
These are expected to be inconsequential finite-size artifacts, since similar features in the nearest-neighbor Kitaev-Heisenberg vanish on larger clusters accessible by density-matrix renormalization group (DMRG) methods\cite{jiang2019field}.

Our results indicate that the IP in \rucl\ 
is only present for strictly \textit{out-of-plane} [111] fields.
This is likely a separate phase that 
 is not necessarily linked to the
 various unconventional experimental observations
 in \rucl\ 
for fields tilted significantly \textit{away} from $[111]$ and
\textit{in-plane} fields
that have been interpreted in terms of a field-induced 
quantum spin liquid\cite{baek2017observation,wang2017magnetic,zheng2017gapless,banerjee2018excitations,kasahara2018half}.

\section{Summary}
In this work, we performed a detailed numerical study  on
 the nature and dynamical response of  field-induced phases in a family of Kitaev-based models
 related by hidden symmetries and duality
transformations. Via exact diagonalization  we investigated
 dynamical spin-structure factors---relevant in INS and ESR
experiments---, dynamical bond correlations---relevant for RIXS and Raman scattering measurements---as well as
 dynamical flux-flux correlations.

Our results indicate that in both FM and AFM Kitaev models, the
first transition where the KSL is suppressed occurs at sufficiently low fields
that the dynamical timescale of the fluxes is likely to be small compared to
the timescales associated with the matter fermion dynamics. As a result, the
phase transition is essentially a large increase of the local flux density. The
fluxes appear to experience effective interactions that are consistent with the
expected features of $c$-fermion-mediated couplings.

We further find that the emergence of a field-induced {\it intermediate phase} (IP) 
 in the AFM Kitaev model is deeply connected to the separation of energy scales
between $K$ and the two-flux gap $\Delta_\text{f}$ and this IP phase appears in the
 Kitaev model for an extensive number of general non-collinear fields for both FM and AFM signs of the interaction.
However,
this phase is remarkably unstable against certain perturbations,
 specifically, a magnetic field with $\mathbf q=\text X$ momenta.
 This may provide clues to its identity.
We also detected additional field-induced
phases ($\Psi_1$) emerging from specific combinations of perturbations, 
 which might be of future interest.

Finally, we demonstrated that
analogues of the AFM KSL and field-induced IP  phase can be found, in principle, 
for models with additional interactions at hidden Kitaev points under uniform fields.
In particular, 
 the IP extends in a finite region of interaction parameter space for out-of-plane fields ($\mathbf h \parallel [111]$),
 such that it can be induced in models with zigzag order, including the \textit{ab-initio}-guided model for \rucl.
For \rucl\ our results predict that the IP 
is present for strictly \textit{out-of-plane} [111] fields. This IP phase is therefore not
 likely linked to the putative field-induced phase recently reported 
in \rucl\ for fields tilted significantly away from $[111]$.

\section*{Acknowledgements}
We are grateful to Radu Coldea, Ciarán Hickey, Simon Trebst and Jeffrey G.\ Rau for stimulating discussions and
acknowledge support by the Deutsche Forschungsgemeinschaft (DFG) through
grant VA117/15-1. Computer time was allotted
at the Centre for Scientific Computing (CSC) in Frankfurt.

\appendix
\crefalias{section}{appsec}
\section{Symmetries and Dualities at Zero and Finite Fields\label{sec:symmetries}}
\begin{figure}
 \includegraphics[width=\linewidth]{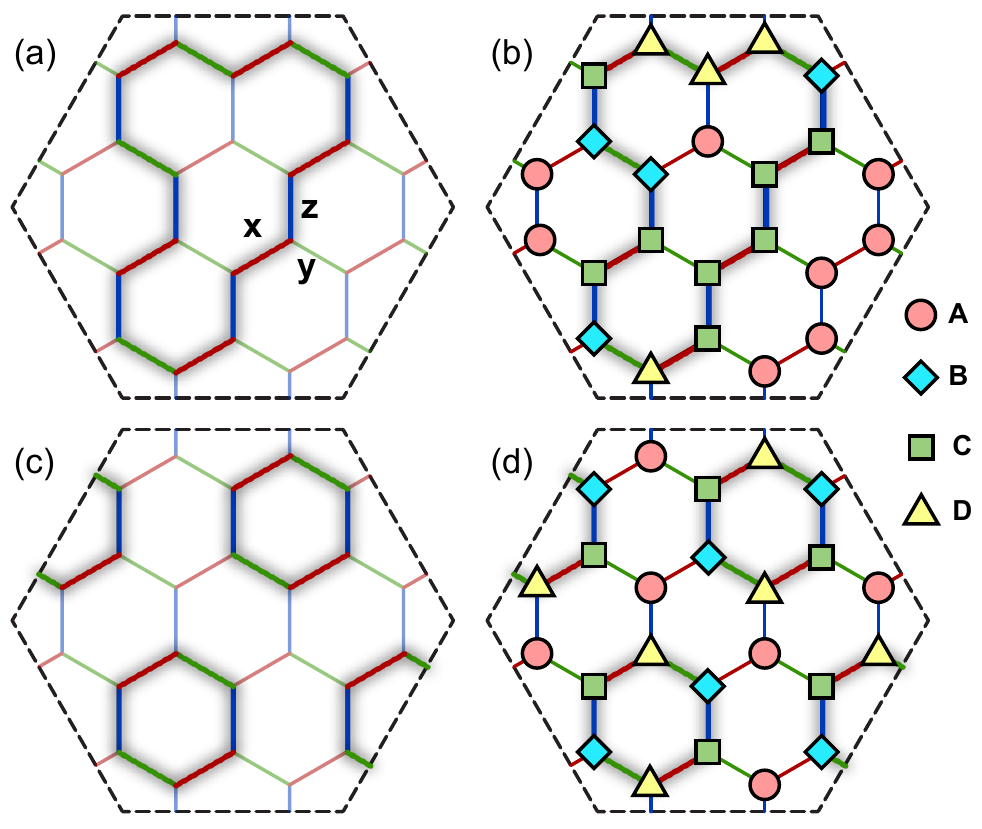}
\caption{Construction of general four-sublattice symmetry transformations for the Kitaev model with $\alpha=\pi$. (a) Example of a loop $L$. The associated operator $W_L$ generates the symmetry transformation shown in (b). The specific operator $W_\text{Klein}$ depicted in (c) leads to the “Klein” transformation (d) employed in Refs.~\onlinecite{chaloupka2010kitaev,chaloupka2013zigzag,chaloupka2015hidden}.}
\label{fig:transforms1}
\end{figure}

In this appendix, we review the construction of general symmetries of the Kitaev model.
 An important feature of the Kitaev Hamiltonian at zero field is the existence of a macroscopic number of mutually commuting Wilson loop operators $\{W_L\}$ that also commute with ${H}_K$. On a torus, any combination of loops $\{L\}$ can be constructed as products of plaquette operators $\{{W}_p\}$ and large loop operators running around the periodic boundaries, 
 i.e.\ ${W}_L = {W}_{p_1}{W}_{p_2}...{W}_{p_n}$. Together, such loop operators form the generators for \textit{local} symmetry transformations $\e^{\frac\imag 2 \alpha W_L}{H}_K \e^{-\frac\imag 2 \alpha W_L}= H_K$ with $0\le\alpha < 2\pi$. For a Hamiltonian ${H} = {H}_K + H_1$, all perturbations   ${H_1}'= \e^{\frac\imag 2 \alpha W_L}{H_1}_K \e^{-\frac\imag 2 \alpha W_L}$ are formally equivalent to $H_1$ for all $L$ and $\alpha$. 
For our purpose, we focus on the specific case $\alpha=\pi$, for which spin operators $\mathcal O = S_i^\mu S_j^\nu S_k^\rho \dots$ transform as $\mathcal O'=\pm \mathcal O$, depending on whether they commute ($+$) or anticommute ($-$) with ${W}_L$. 
 The honeycomb lattice is thus divided into four sublattices, according to the transformation associated with each site:
\begin{align}
W_L: (x^\prime, y^\prime, z^\prime) = \left\{\begin{array}{cc}(x,y,z), & \text{sublattice A} \\ (x,-y,-z),&\text{sublattice B}\\(-x,y,-z),&\text{sublattice C}\\(-x,-y,z),&\text{sublattice D} \end{array} \right.
\label{eq:T_sublattices}
\end{align}
Transformations generated by specific loop operators ${W}_L$ are shown in \cref{fig:transforms1} (a-d). 
A specific example, shown in Fig.~\ref{fig:transforms1}(d), is the 
“Klein” transformation employed in
Refs.~\onlinecite{chaloupka2010kitaev,chaloupka2013zigzag,chaloupka2015hidden},
which is associated to the operator $W_\text{Klein}$ given by the product of plaquette
operators on $\tfrac14$ of the hexagonal plaquettes
(Fig.~\ref{fig:transforms1}(c)).  As noted above, the pure Kitaev Hamiltonian
is mapped to itself by all such transformations, preserving the sign of the
coupling ($K^\prime = K$). However, the operators associated with a magnetic
field generally do not commute with $W_L$. Let us consider a general field 
term described by:
\begin{align}
\fieldterm= -\sum_i h_i^x S_i^x + h_i^y S_i^y + h_i^z S_i^z
\end{align}
where $h_i^\mu$ may take different values at each site. The equivalency of all combinations of local fields $W_L\fieldterm W_L$ generated by different $L$ has several consequences.

In \cref{sec:Kitaev_uni_field}, we discussed the appearance of the gapless IP in the AFM Kitaev model for uniform [111] fields $\fieldterm=H_{[111]}$, where $h_i^x = h_i^y=h_i^z=h$. By symmetry, equivalent states are also induced by all non-uniform field configurations described by $W_LH_{[111]}W_L$. For all such fields, the product of $\gamma$-components of the local fields on a $\gamma$-bond is positive, i.e. $h_i^\gamma h_j^\gamma >0$. 
Since all IP states appearing for $H = H_K + W_LH_{[111]} W_L$ are smoothly connected by continuous unitary transformations, they belong to a common intermediate-field phase. However these transformations do not commute with the spin operators, so that the dynamical spin-structure factor $\mathcal S(\mathbf k, \omega)$ does not provide a unique characterization of the common IP. In contrast, $S_i^\gamma S_j^\gamma$ and $W_p$ commute with all $W_L$, and therefore $\RIXS(\mathbf k,\omega)$ and $\mathcal W(\mathbf k,\omega)$ reflect more intrinsic characteristics that are common across all equivalent IP states.

A similar approach can be used to establish correspondence between the 
FM and AFM Kitaev  models. We define $P_3$, which permutes the spin components $(x,y,z)\to (y,z,x)$ on every site. Combining this with the operator $W_\text{Klein}$ (see \cref{fig:transforms1}(c)) leads to $\mathcal{G} \equiv \e^{\frac{\imag}{2}\pi P_3^{-1} W_\text{Klein} P_3}$, that 
commutes with all $ W_p$, but anticommutes with all bond operators $S_i^\gamma S_j^\gamma$. As a result, $
\mathcal{G}H_K\mathcal{G}^{-1}=-H_K$, providing a duality transformation that relates the AFM and FM Kitaev models at zero field (i.e $K^\prime = -K$). The transformation in real space is:
\begin{align}
\mathcal{G}: (x^\prime, y^\prime, z^\prime) = \left\{\begin{array}{cc}(x,y,z), & \text{sublattice A} \\ (-x,y,-z),&\text{sublattice B}\\(-x,-y,z),&\text{sublattice C}\\(x,-y,-z),&\text{sublattice D} \end{array} \right.
\end{align}
with the sublattice pattern shown in \cref{fig:transforms1}(d). Under the transformation $\mathcal{G}$, a uniform [111]-field $H_{[111]}$ becomes a four-sublattice “tetrahedral” field $\mathcal G H_{[111]} \mathcal G^{-1}$, with local fields $\mathbf h_i$ oriented along the [111], $[\bar{1}1\bar{1}]$, $[1\bar{1}\bar{1}]$, or $[\bar{1}\bar{1}1]$ directions as shown in \cref{fig:tetrahedral_3d}.

If this tetrahedral field is applied to the AFM Kitaev model, it acts as a uniform field in the FM model; the tetrahedral field directly couples to one of the classically degenerate states and opens a gap immediately after the suppression of the KSL. 
 Conversely, if the tetrahedral field is applied to the FM Kitaev model, it
acts as a uniform field in the AFM model, yielding a gapless intermediate
phase. This IP in the FM Kitaev model is continuously connected
 to the IP of the AFM Kitaev
model via the unitary transformation $\mathcal{G}$, and therefore represents
the same phase. 
For field configurations that are dual to one another, the FM 
model displays precisely identical dynamical flux-flux and bond-bond
correlations to those observed in the AFM model. This follows from the fact
that $\mathcal{G}$ commutes with $W_p$ and anticommutes with $S_i^\gamma
S_j^\gamma$.
 
 We remark that for all (uniform and non-uniform) fields considered here, fields that suppress the IP in the FM ($K<0$) or the AFM model ($K>0$) satisfy $\text{sgn}(h_i^\gamma h_j^\gamma) \neq \text{sgn}(K)$ on all $\gamma$-bonds, while conversely fields that induce the IP satisfy on all $\gamma$-bonds 
\begin{equation}
   \text{sgn}(h_i^\gamma h_j^\gamma) = \text{sgn}(K). \label{eq:necessary_condition_app}
\end{equation} 
\Cref{eq:necessary_condition_app} therefore appears to pose a \textit{necessary} condition for the stability of the IP in both Kitaev models.  It leads to opposing signs of nearest-neighbor spin-spin correlations $\braket{S_i^\gamma S_j^\gamma}$ between the low- and high-field limits, and thus provides an additional energy scale where the correlations reverse sign, as seen in \cref{fig:AFM_DSF}(d).

\bibliography{extracted} 

\end{document}